\documentclass[preprint,
 amsmath,amssymb,
 superscriptaddress,
]{revtex4-1}

\usepackage{calc,pifont}
\usepackage{graphicx}
\usepackage{dcolumn}
\usepackage{bm}
\usepackage{lineno}

\begin{document}

\thanks{A footnote to the article title}%
\title{High-Efficiency Second Harmonic Generation of Low-Temporal-Coherent Light Pulse}

\author{Lailin Ji}
\thanks{These authors contributed equally to this Letter.}
\affiliation{Shanghai Institute of Laser Plasma, China Academy of Engineering Physics, 1129 Chenjiashan Road, Shanghai 201800, China}

\author{Xiaohui Zhao}
\thanks{These authors contributed equally to this Letter.}
\affiliation{Shanghai Institute of Laser Plasma, China Academy of Engineering Physics, 1129 Chenjiashan Road, Shanghai 201800, China}

\author{Dong Liu}
\thanks{These authors contributed equally to this Letter.}
\affiliation{Shanghai Institute of Laser Plasma, China Academy of Engineering Physics, 1129 Chenjiashan Road, Shanghai 201800, China}

\author{Yanqi Gao}
\email{liufenggyq@siom.ac.cn}
\affiliation{Shanghai Institute of Laser Plasma, China Academy of Engineering Physics, 1129 Chenjiashan Road, Shanghai 201800, China}
\author{Yong Cui}
\author{Daxing Rao}
\author{Wei Feng}
\author{Fujian Li}
\author{Haitao Shi}
\author{Jiani Liu}
\author{Xiaoli Li}
\author{Lan Xia}
\author{Tao Wang}
\author{Jia Liu}
\author{Pengyuan Du}
\affiliation{Shanghai Institute of Laser Plasma, China Academy of Engineering Physics, 1129 Chenjiashan Road, Shanghai 201800, China}

\author{Xun Sun}
\affiliation{State Key Laboratory of Crystal Materials, Shandong University, Jinan, 250100, China}

\author{Weixin Ma}
\author{Zhan Sui}
\affiliation{Shanghai Institute of Laser Plasma, China Academy of Engineering Physics, 1129 Chenjiashan Road, Shanghai 201800, China}

\author{Xianfeng Chen}
\affiliation{School of Physics and Astronomy, Shanghai Jiao Tong University, 800 Dongchuan Road, Shanghai 200240, China}

\date{\today}

\begin{abstract}

The nonlinear frequency conversion of low-temporal-coherent light holds a variety of applications and has attracted considerable interest. However, its physical mechanism remains relatively unexplored, and the conversion efficiency and bandwidth are extremely insufficient. Here, considering the instantaneous broadband characteristic, we establish a model of second harmonic generation (SHG) of low-temporal-coherent pulse, and reveal its differences from the coherent conditions. It is found that the second harmonic (SH) of low-temporal-coherent light is produced by not only the degenerate SH processes but also crossed sum-frequency processes. On the basis of this, we propose a method for realizing low-temporal-coherent SHG with high efficiency and broad bandwidth, and experimentally demonstrate a conversion efficiency up to 70\% with a bandwidth of 3.1 THz (2.9 nm centered at 528 nm). To the best of our knowledge, this is the highest efficiency and broadest bandwidth of low-temporal-coherent SHG, and its efficiency is almost the same with that of the narrowband coherent condition. Furthermore, the spectral evolution characteristics of the broadband low-temporal-coherent pulse in SHG process are revealed in experiments, that the SH power spectral density (PSD) is proportional to the self-convolution of the fundamental wave PSD, which is greatly different from that of the coherent process. Our research opens a door for the study of the low-coherent nonlinear optical processes.

\end{abstract}

\maketitle

One of the most challenging issues in laser driven inertial confinement fusion (ICF) is the suppression of laser-plasma instabilities (LPI) when intense laser transmits through a surrounding plasma \cite{Betti2016Inertial,Hurricane2014Fuel,Lindl2014Review}. The key to solving this problem is decoherence of high coherent diver laser to reduce instabilities caused by nonlinear processes \cite{Labaune2007Laser,Glenzer2007Experiments}, such as self-focusing, stimulated Raman scattering (SRS),  stimulated Brillouin scattering (SBS) and crossed-beam energy transfer (CBET) \cite{Lindl2004The,Michel2009Tuning,Moody2012Multistep}. At present, beam smoothing techniques, which reduce the temporal coherence (such as SSD, PS)  \cite{Skupsky1989Improved,Dorrer:14,Zhong:18,Rothenberg2000Polarization} and spatial coherence (such as RPP, CPP, ISI, LA) \cite{PhysRevLett.53.1057,Lin:95,PhysRevLett.56.2807,Deng:86} of laser, are the main solution and have been widely used in laser-fusion facilities around the world \cite{Spaeth2016National,Dorrer2017Spectrally,FLEUROT2005147,Jiang2018Experimental}. In general, the bandwidth of laser for ICF imposed by modulator is about 100 GHz, loosely speaking, an instantaneous frequency that varies periodically in tim \cite{Haynam:07}. These factors severely limit the smoothing speed and smoothing effect of laser beam. Plenty of experiments and analyses indicate that, the adverse effects of LPI under fusion conditions are difficult to effectively overcome using current beam smoothing techniques \cite{doi:10.1063/1.4946016,PhysRevLett.120.055001,PhysRevLett.115.055003}. A more straightforward solution is using broadband low-temporal-coherent laser sources ($\Delta\nu/\nu>1\%$) for ignition \cite{Labaune2007Laser,Eimerl2014StarDriver,PhysRevLett.70.2738,PhysRevE.97.061202}.  These laser sources have much broader bandwidth, more plenty spectrum components, and much lower coherence for achieving a better smoothing effect. This method is expected to alleviate the LPI problem that has plagued the fusion field for many years. For the LPI, shorter laser wavelengths can also improve the coupling efficiency of the laser-plasma and reduce harmful processes. Currently, most Nd:glass laser fusion facilities operate at the third harmonic (351 nm). However, laser damages caused by ultraviolet severely limit the facility output ability and greatly increase the operation cost \cite{031b69f3114540a4811dc0cee2ed0b87}.  In addition, the narrow acceptance bandwidth of existing third harmonic generation method restricts the effect of the current beam smoothing technologies. The low-coherent second harmonic driver will greatly alleviate or even solve the above problems \cite{Chambonneau:18}. Hence, frequency conversion techniques for broadband low-temporal-coherent laser may open the door to use the second-harmonic (SH) laser for ignition \cite{doi:10.1063/1.1687725,Heestand:08}.

Frequency conversion is a fundamental issue in nonlinear optics. The typical applications are the second harmonic generation and sum-frequency generation of the narrow-band coherent laser in ICF, which have a conversion efficiency up to 80\%. With the development of ultrashort pulses, frequency conversion of broadband coherent pulses (chirped pulse and compressed pulse) has been widely studied. The greatest difficulty is satisfying the phase matching (PM) condition and the group-velocity matching (GVM) condition simultaneously, which determine the conversion efficiency and spectral bandwidth of nonlinear process \cite{1081948}. The multi-crystal scheme \cite{1073521}, angular spectral dispersion (ASD) method \cite{135282,Nakatsuka1993Partially} and partially deuterated KDP crystal (DKDP) \cite{Webb:92} are respectively developed for second harmonic generation (SHG) of broadband coherent pulses. The highest SH conversion efficiency ($\eta$) has been achieved is 75\% under an extremely high intensity about 380 GW/cm$^2$ \cite{Hillier:13}, and the bandwidth ($\Delta\lambda$) about 3 nm is achieved. For broadband coherent pulses, achieving high-efficiency harmonic conversion under normal condition($\sim$GW/cm$^2$) is still a challenge. Recently, SHG of low-coherent pulse has attracted considerable interest. Some exploration researches about the theory \cite{PhysRevA.36.202,Cai:07} and experiment \cite{Dmitriev_2012} on SHG of low-spatial-coherent pulses are carried out, and the conversion efficiency has been achieved is about 35\% with divergence of 3.5 mrad \cite{Vasin2013Second}. However, the researches for SHG of low-temporal-coherent pulses are relatively insufficient, and the physical mechanism is still not conclusive. There are three primary questions waiting to be answered: (1) What is the essential difference between the low-temporal-coherence frequency conversion processes and the coherent processes? (2) How to achieve a high-efficiency low-temporal-coherence frequency conversion? (3) What are the coherence characteristics of the generated harmonic wave?

In this Letter, we study the physical mechanism of the low-temporal-coherent SHG process based on the instantaneous broadband characteristic, firstly. The difference from the coherent processes is clarified from the perspective of statistical optics, and the distribution of SH power spectral density (PSD) is theoretically predicted (that it is a self-convolution of the fundamental wave PSD). Furthermore, we propose and experimentally demonstrate a method for realizing low-temporal-coherent broadband SHG with a conversion efficiency up to 70\%. The characteristics of SH wave and fundamental wave (FW) are compared and analyzed. Through the analysis of spectral evolution characteristics, which is consistent with the theoretical prediction by the model proposed in this paper, the physical mechanism is further verified. Our research has great significance for the study of low-temporal-coherence nonlinear optical processes.

\section*{Results}
\textbf{Statistical analyses of the SHG process of low-temporal-coherent pulse.} For simplicity, the low-temporal-coherent pulses discussed here, unless otherwise specified, are spatial coherent. Generally, the low-temporal-coherent light pulse has the characteristic of instantaneous broadband. Different from compressed pulse [Fig. 1(a)], chirped pulse [Fig. 1(b)] and modulated pulse [Fig.1(c)], the frequency component of instantaneous broadband pulse [Fig. 1(d)] has a comprehensive distribution at any time within the pulse duration, resulting in a more complicated process. Under the precondition of satisfying the conditions of the PM and the GVM simultaneously, multiple physical processes will generate in nonlinear medium, since the instantaneous broadband pulse has multiple frequency components at the same time. Therefore, we propose the schematic of instantaneous broadband SHG, which is shown in Fig. 1(e): (1), (3) and (4) are the degenerate second harmonic processes at different frequency; (2) denotes the sum-frequency process of two FW photons with different energy. It can be inferred that, SH photons with the same wavelength are generated by both degenerate second harmonic process and different sum-frequency processes. The distribution of SH spectrum is determined by the spectrum and statistical characteristics of FW.

\begin{figure}[htp!]
\centering\includegraphics[width=8.5 cm]{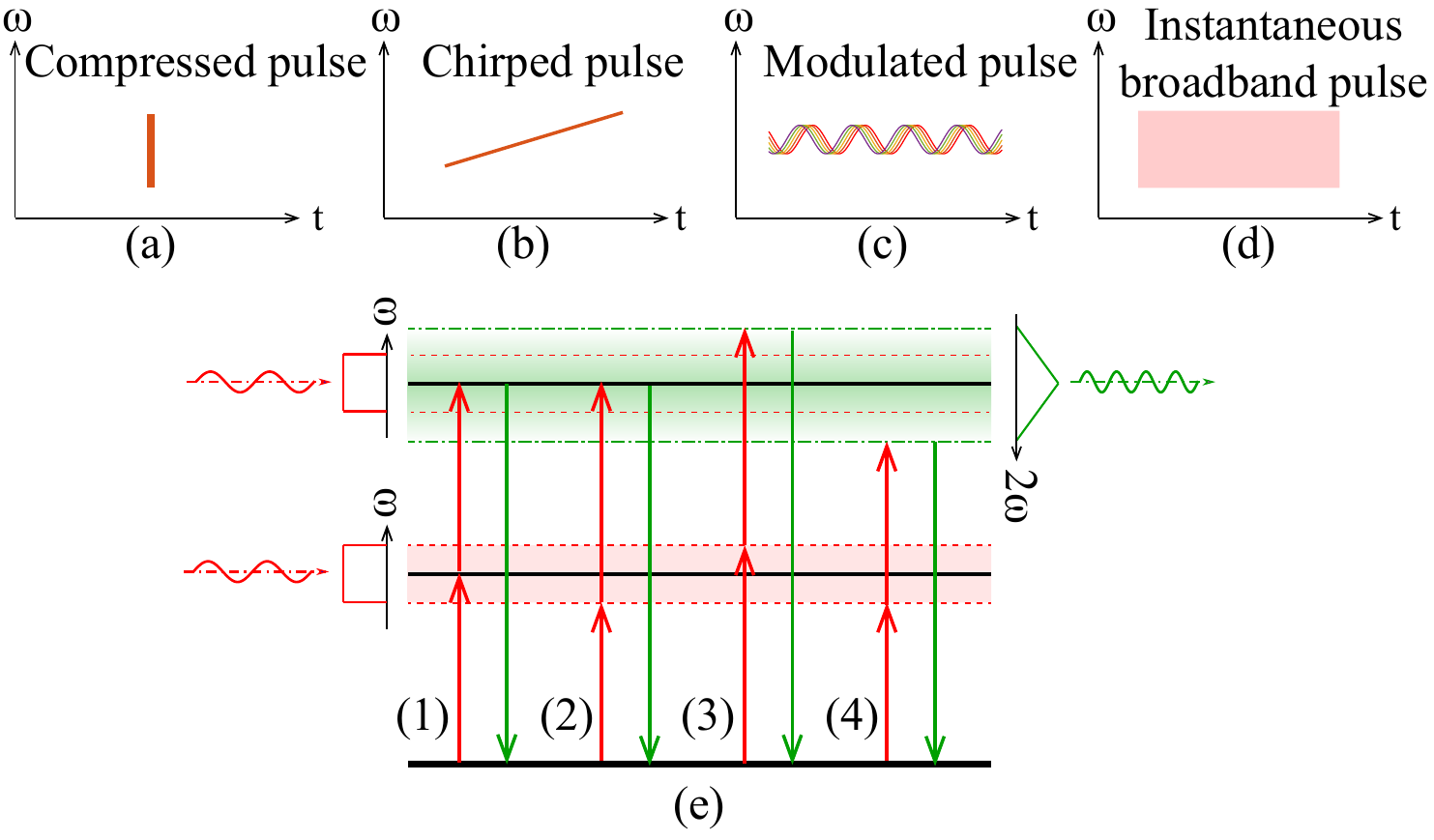}
\caption{The spectrum distributions of (a) compressed pulse; (b) chirped pulse; (c) modulated pulse and (d) instantaneous broadband pulse. (e) Schematic of instantaneous broadband SHG.}
\end{figure}

To analyze the SHG process of low-temporal-coherent pulse, statistical optics is introduced to calculate the intensity of SH wave. Under the condition of perfect matching, the complex amplitude of the SH can be calculated as $E_{2\omega}=\sqrt{\eta}E_{\omega}^2/\lvert E_{\omega}\rvert$, where $E_{\omega}$ is the complex amplitude of the FW and $\eta$ is the conversion efficiency. So the temporal autocorrelation function of SH is:
\begin{equation}
\gamma_{2\omega}(\tau)=\langle E_{2\omega}^{\ast}\left(t\right)E_{2\omega}\left(t^{\prime}\right)\rangle\propto\langle E_{\omega}^{2\ast}\left(t\right)E_{\omega}^2\left(t^{\prime}\right)\rangle,
\end{equation}
where $t$ and $t^{\prime}$ are two moments within the pulse duration, $\tau\equiv t^{\prime}-t$ and the symbol $\langle\cdot\rangle$ represents an average for infinite time. According to the Wiener-Khinchin theorem, the power spectral density (PSD) is the Fourier transform of the temporal autocorrelation function, so that the PSD of SH can be derived from Eq. (1):
\begin{equation}
I_{2\omega}(\nu)=\Im\left\{\gamma_{2\omega}(\tau)\right\}\propto\langle\lvert E_{\omega}(\nu)\otimes E_{\omega}(\nu)\rvert^2\rangle.
\end{equation}
The frequency-domain electric field $E_{\omega}(\nu)=\sqrt{I_{\omega}(\nu)}\exp[i\phi(\nu)]$ is the Fourier transform of $E_{\omega}(t)$. $I_{\omega}(\nu)$ and $\phi(\nu)$ is the PSD and spectral phase of FW, respectively. For different source, $E_{\omega}(\nu)$ has different statistical characteristics resulting in different spectral types. For the instantaneous broadband source, its amplitude and spectral phase are statistically independent, and the phase is evenly distributed in $[-\pi,\pi]$. After a simple derivation, we can get the relationship between the PSD of SH and that of the FW:
\begin{equation}
I_{2\omega}(\nu)\propto \langle \left[\sqrt{I_{\omega}(\nu)}\otimes  \sqrt{I_{\omega}(\nu)}\right]^2\rangle.
\end{equation}
From Eq. (3), the PSD of SH is proportional to the self-convolution of the PSD of FW. It is different from the coherent SHG process, in which the electric field spectrum of SH is proportional to the self-convolution of that of FW.

\textbf{Experimental results and analyses.} For an experimental demonstration, a superluminescent diode (SLD) pulse source amplified by a Nd:phosphate glass rod laser system was chosen as the pumped laser. The time-domain waveform and spectral distribution of this kind of pulse is independent \cite{cui2019High}, which presents instantaneous broadband characteristic as show in Fig. 1(d).  The spatial profile of the beam is nearly a 12th-order super-Gaussian with a size of 42 mm$\times$42 mm. It delivers a intensity up to 0.75 GW/cm$^2$ within a 3 ns pulse duration. The output spectral width is up to 10.2 nm with a center wavelength at 1057 nm. The coherent time is about 318 fs measured in experiments, far less than the pulse duration, demonstrating that the pulse has low temporal coherence. Moreover, the measurements of near-field and far-field profiles indicate that the light source has good spatial coherence. After a nonlinear crystal, three dichroic mirrors (M1-M3) were set to separate the SH wave and the residual FW, as shown in Fig. 2. The energy of SH was measured by an energy meter. The spectrometer behind the M1 is utilized to detect the spectrum of FW and SH. Behind M2, the leaked SH wave was split into two beams, which were focused by lens L1 and L2 for the detection of far-field profile and temporal waveform, respectively.

\begin{figure}[ht!]
\centering\includegraphics[width=8.0 cm]{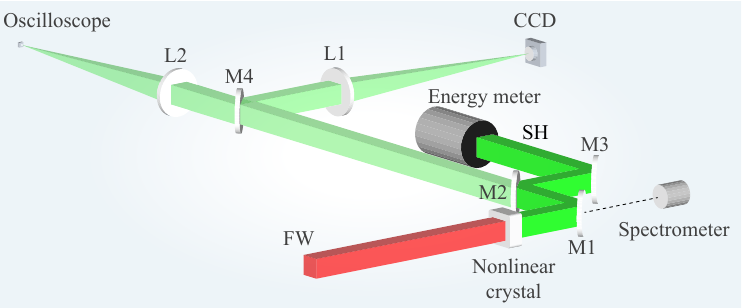}
\caption{Experimental setup: M1-M3 are second-harmonic beam splitters, M4 is a beam splitter, L1 and L2 are focusing lenses with a focal length of 1 m, a CCD and an oscilloscope measure the far-field profile and temporal waveform of SH.}
\end{figure}

For simultaneously satisfying the PM and GVM conditions, a 15\% DKDP crystal with a cutting angle $\theta=41^{\circ}$ for type-\uppercase\expandafter{\romannumeral1} phase-matching was utilized, which center wavelength of phase matching is 1057 nm. Based on the numerical simulation of the nonlinear coupled-wave equations, the crystal length was designed to be 32 mm. It provides an acceptance bandwidth of about 12 nm at the retracing point when the SHG falls into the regime of saturation. The cross-sectional dimension of the crystal is 70 mm$\times$70 mm. 

\begin{figure}[ht!]
\centering\includegraphics[width=8.0cm]{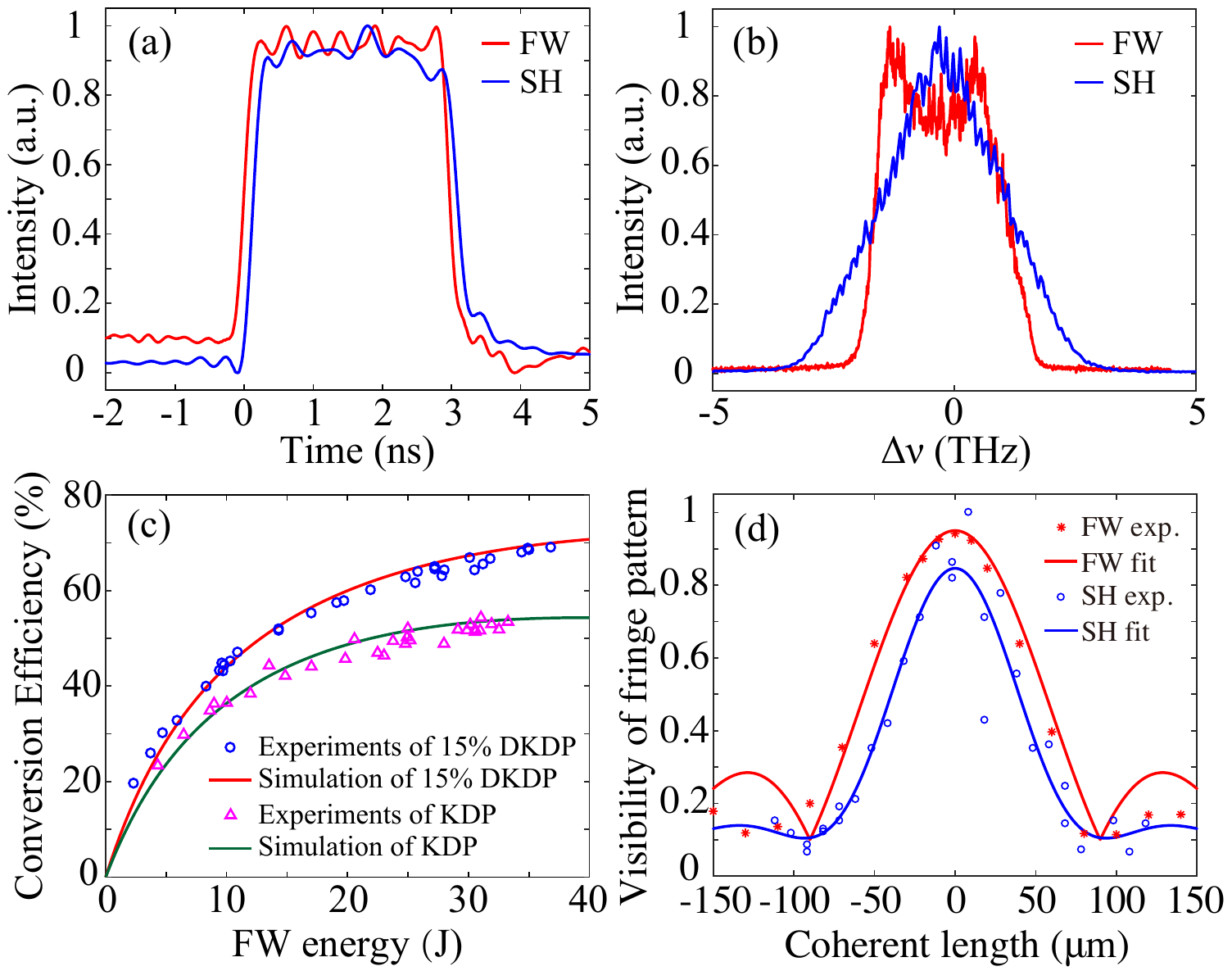}
\caption{Experimental results of SHG: temporal waveforms (a) and the spectrums (b) of FW and SH; (c) the conversion efficiency with 15\% DKDP and KDP crystal as a function of the FW energy; (d) the coherence length of SH.}
\end{figure}

Figure 3 presents the primary experimental results. The FW is a square waveform with a pulse duration of 3 ns [Fig. 3(a)], measured by a 4-GHz oscilloscope. The time waveform of SH is the same with that of FW. The spectrum of FW [Fig. 3(b)] has an approximately rectangular distribution. The full width at half maxima (FWHM) of the spectrum is 10.2 nm (2.7 THz). The spectrum of SH has a triangular profile with a bandwidth (FWHM) of 2.9 nm (3.1 THz). To the best of our knowledge, this is the broadest bandwidth of low-temporal-coherent SH pulse. In addition, the spectrum distribution in Fig. 3(b) shows that the spectrum of SH proportional to the self-convolution of that of the FW, which is in well agreement with our theoretical prediction indicated above.

Figure 3(c) shows the relationship of the SH  conversion efficiency versus FW energy. The conversion efficiency was calculated by dividing the SH energy by the incident FW energy (sampling measured before the nonlinear crystal, not shown in Fig. 2) with compensated for the energy lost by the three dichroic mirrors ($\sim3\%$). In our experiments, the highest conversion efficiency achieved by 15\% DKDP crystal was 70\%. To the best of our knowledge, this is the highest conversion efficiency of low-temporal-coherent SHG. For comparison, a type-\uppercase\expandafter{\romannumeral1} phase-matching KDP crystal  ($\theta^{\prime}=40^{\circ}$) with the same size of 15\% DKDP was utilized to repeat the experiments. The acceptance bandwidth of KDP crystal is 2 nm. The highest conversion efficiency of it is about 55\%, significantly smaller than that of 15\% DKDP crystal. The curves in Fig. 3(c) are the simulation results based on the nonlinear coupled equation considering the statistical characteristics of the instantaneous broadband source. Simulation and experimental results are well in agreement with each other. Utilizing  the 15\% DKDP crystal, the conversion efficiency was 70\% at the FW intensity density of 0.75 GW$/$cm$^2$. Theoretically, at the conventional power density of high power laser facility for ICF (3$\sim$4 GW$/$cm$^2$), it can achieve a conversion efficiency about 80\%, which is almost the same with the highest conversion efficiency on narrowband coherent laser facility. The conversion efficiency is a pivotal parameter for some applications, especially ICF, which decides the highest available driver energy.

Experimentally, the temporal coherence length of the FW and SH was measured by a Michelson interferometer, and the result was shown in Fig. 3(d). Theoretically, the contrast function $\Gamma(\tau)$ is the module of the temporal autocorrelation function $\gamma_{2\omega}(\tau)$. From Eq. 2, it can derive that $\Gamma(\tau)=\lvert\Im^{-1}\{I_{2\omega}(\nu)\}\rvert$. The spectrum of SH has a triangular distribution in our experiment, so that the contrast function is in the form of $\text{sinc}^2(\Delta\nu\tau)$, where $\Delta\nu$ is the FWHM of the SH spectrum. Similarly, the contrast function of FW is in the form of $\text{sinc}(\Delta\nu_1\tau)$, where $\Delta\nu_1$ is the FWHM of the FW spectrum. The relationship of the visibility of fringe pattern versus the optical path difference is well in agreement with the theoretical prediction. The coherence time of SH is 300 fs, which is approximate to that of the FW (318 fs). It is far less than the pulse duration shown in Fig. 3(a), demonstrating that the SHG process has little effect on the low coherent characteristic.

\begin{figure}[ht!]
\centering\includegraphics[width=8.5 cm]{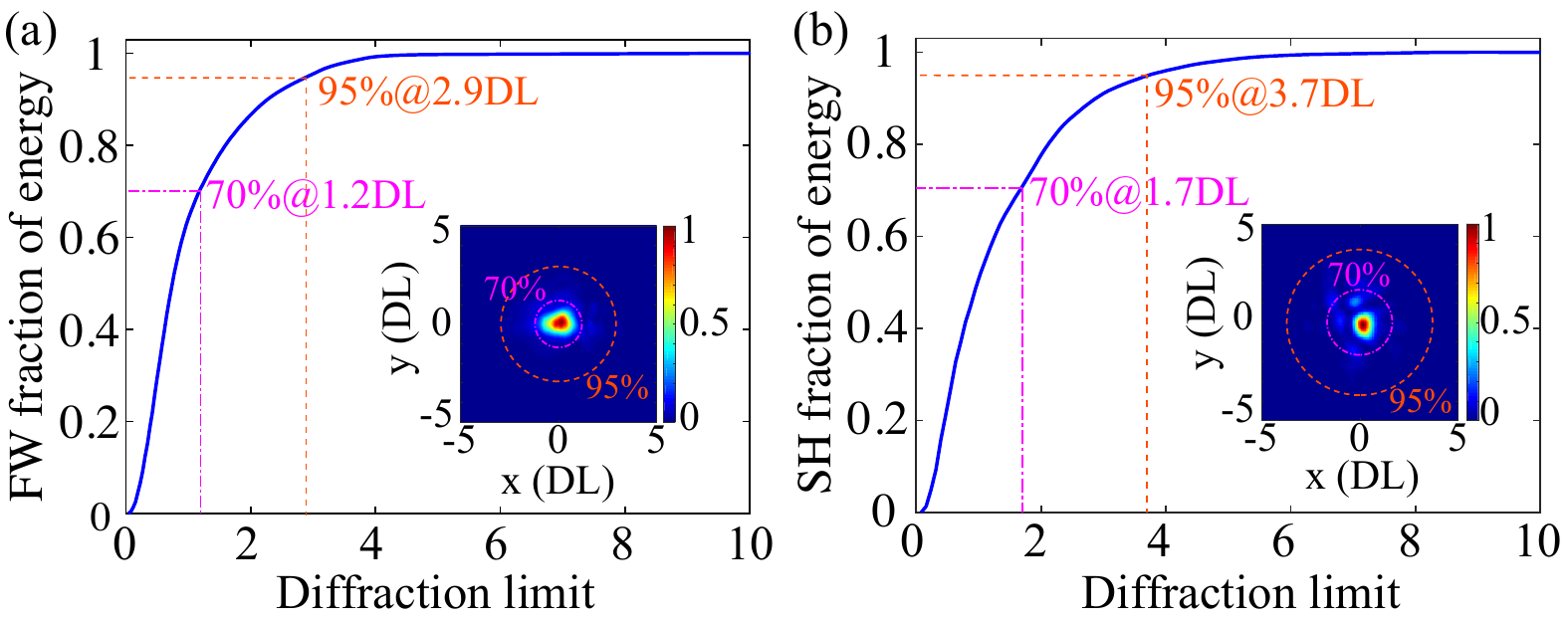}
\caption{Energy concentration rate curve of FW (a) and SH (b) focusing by lenes with a focal length of 1.026 m.  The insets are the far-field profile of FW and SH, respectively.}
\end{figure}

Far field focusing characteristics of FW and SH are shown in Fig. 4. For SH wave, more than 70\% of the energy is in the range of 1.7 times the diffraction limit (DL), and more than 95\% energy is in the range of 3.7 times DL, which is slightly degraded compared to that of the FW. It shows that, the low-temporal-coherent pulse utilized in experiments is spatial coherent, satisfying assumptions of the model in this paper. After the conversion process, the SH pulse holds an excellent far-field performance, which presents a good focusing capability required by a lot of applications.

\textbf{The evolutions of SH spectrum.} To further verify our theoretical inference on the physical mechanism of the low-temporal-coherent nonlinear frequency conversion processes, the evolutions of spectral characteristics were investigated. We designed the spectrum distribution of FW with a bimodal structure with center wavelengths of 1052 nm and 1060 nm, respectively, as shown in Fig. 5(a). Also, its waveform was square in the time domain. The theoretical self-convolution of FW spectrum is shown in Fig. 5(b), which has three peaks with center wavelengths of 526 nm, 528 nm and 530 nm. Figure 5(c) and (d) present the evolutions of SH spectrum varying with incident angle $\theta$ in the 15\% DKDP crystal and KDP crystal, respectively, where $\theta$ is defined as the external rotation angle with the o-axis of the crystal. 

\begin{figure}[ht!]
\centering\includegraphics[width=8.5 cm]{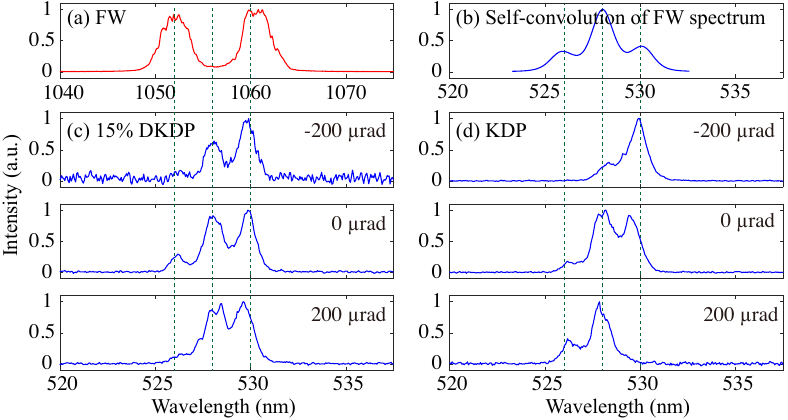}
\caption{Evolution of SH spectrum varying with the incident angles at nonlinear crystals: (a) modulated spectrum of FW; (b) self-convolution of FW spectrum; (c) SH spectrum of the 15\% DKDP crystal and (d) the KDP crystal at different angles.}
\end{figure}

At the angle of $0\ \mu\text{rad}$, corresponding to the retracing-point of phase-matching, the spectrum of SH has similar shape with the theoretically prediction in Fig. 5(b), which confirms the self-convolution relationship described in Eq. (3). Although the frequency components of FW are lacking around 1056 nm, the SH has a spectrum peak at 528 nm. It verified that in the low-temporal-coherent SHG process, the harmonic waves are produced by not only the degenerate second harmonic process but also crossed sum-frequency processes. The physical mechanism proposed in this paper is demonstrated.

Moreover, in experiment, the acceptance bandwidth provided by a nonlinear crystal is not infinite, which is equivalent to a filtering process. We verified this filtering process by adjusting the crystal angles to change the center wavelength of the phase matching. Since the acceptance bandwidth of the KDP crystal is narrow, the filtering effect is more obvious than that of the 15\% DKDP crystal. The phase matching angle is monotonic for wavelength. Varying with the incident angles, the peak of SH spectrum is shifted. The above experimental results not only demonstrate our theoretical analysis, but also predict a broader-band SHG by modulating the spectrum of fundamental wave.

\section*{Discussion}
We revealed the novel physical mechanism of low-temporal-coherent SHG process based on the instantaneous broadband characteristic. The essential difference with harmonic processes of coherent light is that the spectral components of SH are produced by both the degenerate SHG and the crossed sum-frequency processes, and the convolution relationship is between the PSD of SH and FW, not the electric field spectrum. The method of high-efficiency low-temporal-coherent SHG with broadband was proposed. The conversion efficiency in our experiment was up to 70\% (at 0.75 GW$/$cm$^2$), and the bandwidth is 3.1 THz (2.9 nm). The low-temporal coherence characteristric was kept during the nonlinear process. Moreover, the PSD evaluation relationship during the SHG process was demonstrated experimentally, which is consistent with theoretical prediction. The analyses in this paper will also applicable to third- and even higher-order harmonic conversion processes of low-temporal-coherent light. Our research will have a further impact on the study of nonlinear optical process of low-coherence light.\\

\section*{Acknowledgements}
This work is supported in part by the Science Challenge Project (Grant No. TZ2016005), the National Natural Science Foundation of China under Grant Nos. 11804321, 11604317 and 11604318.

\section*{Author contributions}
Lailin Ji conceived the original idea and model simulation; Dong Liu?Lailin Ji and Xiaohui Zhao carried out the experiments; Xun Sun provided key experimental material; Xiaohui Zhao, Lailin Ji, Yanqi Gao and Xianfeng Chen produced the manuscript and interpreted the results; Yanqi Gao, Yong Cui, Daxing Rao, Wei Feng, Fujian Li, Haitao Shi, Jiani Liu, Xiaoli Li, Lan Xia, Tao Wang, Jia Liu, Pengyuan Du contributed to experiments and data analyses; Yanqi Gao, Weixin Ma and Zhan Sui supervised the project. All authors participated in discussions and reviewed the manuscript.


\begin{thebibliography}{40}%
\makeatletter
\providecommand \@ifxundefined [1]{%
 \@ifx{#1\undefined}
}%
\providecommand \@ifnum [1]{%
 \ifnum #1\expandafter \@firstoftwo
 \else \expandafter \@secondoftwo
 \fi
}%
\providecommand \@ifx [1]{%
 \ifx #1\expandafter \@firstoftwo
 \else \expandafter \@secondoftwo
 \fi
}%
\providecommand \natexlab [1]{#1}%
\providecommand \enquote  [1]{``#1''}%
\providecommand \bibnamefont  [1]{#1}%
\providecommand \bibfnamefont [1]{#1}%
\providecommand \citenamefont [1]{#1}%
\providecommand \href@noop [0]{\@secondoftwo}%
\providecommand \href [0]{\begingroup \@sanitize@url \@href}%
\providecommand \@href[1]{\@@startlink{#1}\@@href}%
\providecommand \@@href[1]{\endgroup#1\@@endlink}%
\providecommand \@sanitize@url [0]{\catcode `\\12\catcode `\$12\catcode
  `\&12\catcode `\#12\catcode `\^12\catcode `\_12\catcode `\%12\relax}%
\providecommand \@@startlink[1]{}%
\providecommand \@@endlink[0]{}%
\providecommand \url  [0]{\begingroup\@sanitize@url \@url }%
\providecommand \@url [1]{\endgroup\@href {#1}{\urlprefix }}%
\providecommand \urlprefix  [0]{URL }%
\providecommand \Eprint [0]{\href }%
\providecommand \doibase [0]{http://dx.doi.org/}%
\providecommand \selectlanguage [0]{\@gobble}%
\providecommand \bibinfo  [0]{\@secondoftwo}%
\providecommand \bibfield  [0]{\@secondoftwo}%
\providecommand \translation [1]{[#1]}%
\providecommand \BibitemOpen [0]{}%
\providecommand \bibitemStop [0]{}%
\providecommand \bibitemNoStop [0]{.\EOS\space}%
\providecommand \EOS [0]{\spacefactor3000\relax}%
\providecommand \BibitemShut  [1]{\csname bibitem#1\endcsname}%
\let\auto@bib@innerbib\@empty
\bibitem [{\citenamefont {Betti}\ and\ \citenamefont
  {Hurricane}(2016)}]{Betti2016Inertial}%
  \BibitemOpen
  \bibfield  {author} {\bibinfo {author} {\bibfnamefont {R.}~\bibnamefont
  {Betti}}\ and\ \bibinfo {author} {\bibfnamefont {O.~A.}\ \bibnamefont
  {Hurricane}},\ }\href@noop {} {\bibfield  {journal} {\bibinfo  {journal}
  {Nature Physics}\ }\textbf {\bibinfo {volume} {12}},\ \bibinfo {pages} {435}
  (\bibinfo {year} {2016})}\BibitemShut {NoStop}%
\bibitem [{\citenamefont {Hurricane}\ \emph {et~al.}(2014)\citenamefont
  {Hurricane}, \citenamefont {Callahan}, \citenamefont {Casey}, \citenamefont
  {Celliers}, \citenamefont {Cerjan}, \citenamefont {Dewald}, \citenamefont
  {Dittrich}, \citenamefont {D$\text{\"o}$ppner}, \citenamefont {Hinkel},
  \citenamefont {Hopkins}, \citenamefont {Kline}, \citenamefont {LePape},
  \citenamefont {Ma}, \citenamefont {MacPhee}, \citenamefont {Milovich},
  \citenamefont {Pak}, \citenamefont {Park}, \citenamefont {Patel},
  \citenamefont {Remington}, \citenamefont {Salmonson}, \citenamefont
  {Springer},\ and\ \citenamefont {Tommasini}}]{Hurricane2014Fuel}%
  \BibitemOpen
  \bibfield  {author} {\bibinfo {author} {\bibfnamefont {O.~A.}\ \bibnamefont
  {Hurricane}}, \bibinfo {author} {\bibfnamefont {D.~A.}\ \bibnamefont
  {Callahan}}, \bibinfo {author} {\bibfnamefont {D.~T.}\ \bibnamefont {Casey}},
  \bibinfo {author} {\bibfnamefont {P.~M.}\ \bibnamefont {Celliers}}, \bibinfo
  {author} {\bibfnamefont {C.}~\bibnamefont {Cerjan}}, \bibinfo {author}
  {\bibfnamefont {E.~L.}\ \bibnamefont {Dewald}}, \bibinfo {author}
  {\bibfnamefont {T.~R.}\ \bibnamefont {Dittrich}}, \bibinfo {author}
  {\bibfnamefont {T.}~\bibnamefont {D$\text{\"o}$ppner}}, \bibinfo {author}
  {\bibfnamefont {D.~E.}\ \bibnamefont {Hinkel}}, \bibinfo {author}
  {\bibfnamefont {L.~F.~B.}\ \bibnamefont {Hopkins}}, \bibinfo {author}
  {\bibfnamefont {J.~L.}\ \bibnamefont {Kline}}, \bibinfo {author}
  {\bibfnamefont {S.}~\bibnamefont {LePape}}, \bibinfo {author} {\bibfnamefont
  {T.}~\bibnamefont {Ma}}, \bibinfo {author} {\bibfnamefont {A.~G.}\
  \bibnamefont {MacPhee}}, \bibinfo {author} {\bibfnamefont {J.~L.}\
  \bibnamefont {Milovich}}, \bibinfo {author} {\bibfnamefont {A.}~\bibnamefont
  {Pak}}, \bibinfo {author} {\bibfnamefont {H.~S.}\ \bibnamefont {Park}},
  \bibinfo {author} {\bibfnamefont {P.~K.}\ \bibnamefont {Patel}}, \bibinfo
  {author} {\bibfnamefont {B.~A.}\ \bibnamefont {Remington}}, \bibinfo {author}
  {\bibfnamefont {J.~D.}\ \bibnamefont {Salmonson}}, \bibinfo {author}
  {\bibfnamefont {P.~T.}\ \bibnamefont {Springer}}, \ and\ \bibinfo {author}
  {\bibfnamefont {R.}~\bibnamefont {Tommasini}},\ }\href@noop {} {\bibfield
  {journal} {\bibinfo  {journal} {Nature}\ }\textbf {\bibinfo {volume} {506}},\
  \bibinfo {pages} {343} (\bibinfo {year} {2014})}\BibitemShut {NoStop}%
\bibitem [{\citenamefont {Lindl}\ \emph {et~al.}(2014)\citenamefont {Lindl},
  \citenamefont {Landen}, \citenamefont {Edwards},\ and\ \citenamefont
  {Moses}}]{Lindl2014Review}%
  \BibitemOpen
  \bibfield  {author} {\bibinfo {author} {\bibfnamefont {J.}~\bibnamefont
  {Lindl}}, \bibinfo {author} {\bibfnamefont {O.}~\bibnamefont {Landen}},
  \bibinfo {author} {\bibfnamefont {J.}~\bibnamefont {Edwards}}, \ and\
  \bibinfo {author} {\bibfnamefont {E.}~\bibnamefont {Moses}},\ }\href@noop {}
  {\bibfield  {journal} {\bibinfo  {journal} {Physics of Plasmas}\ }\textbf
  {\bibinfo {volume} {21}},\ \bibinfo {pages} {339} (\bibinfo {year}
  {2014})}\BibitemShut {NoStop}%
\bibitem [{\citenamefont {Labaune}(2007)}]{Labaune2007Laser}%
  \BibitemOpen
  \bibfield  {author} {\bibinfo {author} {\bibfnamefont {C.}~\bibnamefont
  {Labaune}},\ }\href@noop {} {\bibfield  {journal} {\bibinfo  {journal}
  {Nature Physics}\ }\textbf {\bibinfo {volume} {3}},\ \bibinfo {pages} {680}
  (\bibinfo {year} {2007})}\BibitemShut {NoStop}%
\bibitem [{\citenamefont {Glenzer}\ \emph {et~al.}(2007)\citenamefont
  {Glenzer}, \citenamefont {Froula}, \citenamefont {Divol}, \citenamefont
  {Dorr}, \citenamefont {Berger}, \citenamefont {Dixit}, \citenamefont
  {Hammel}, \citenamefont {Haynam}, \citenamefont {Hittinger},\ and\
  \citenamefont {Holder}}]{Glenzer2007Experiments}%
  \BibitemOpen
  \bibfield  {author} {\bibinfo {author} {\bibfnamefont {S.~H.}\ \bibnamefont
  {Glenzer}}, \bibinfo {author} {\bibfnamefont {D.~H.}\ \bibnamefont {Froula}},
  \bibinfo {author} {\bibfnamefont {L.}~\bibnamefont {Divol}}, \bibinfo
  {author} {\bibfnamefont {M.}~\bibnamefont {Dorr}}, \bibinfo {author}
  {\bibfnamefont {R.~L.}\ \bibnamefont {Berger}}, \bibinfo {author}
  {\bibfnamefont {S.}~\bibnamefont {Dixit}}, \bibinfo {author} {\bibfnamefont
  {B.~A.}\ \bibnamefont {Hammel}}, \bibinfo {author} {\bibfnamefont
  {C.}~\bibnamefont {Haynam}}, \bibinfo {author} {\bibfnamefont {J.~A.}\
  \bibnamefont {Hittinger}}, \ and\ \bibinfo {author} {\bibfnamefont {J.~P.}\
  \bibnamefont {Holder}},\ }\href@noop {} {\bibfield  {journal} {\bibinfo
  {journal} {Nature Physics}\ }\textbf {\bibinfo {volume} {3}},\ \bibinfo
  {pages} {716} (\bibinfo {year} {2007})}\BibitemShut {NoStop}%
\bibitem [{\citenamefont {Lindl}\ \emph {et~al.}(2004)\citenamefont {Lindl},
  \citenamefont {Amendt}, \citenamefont {Berger}, \citenamefont {Glendinning},
  \citenamefont {Glenzer}, \citenamefont {Haan}, \citenamefont {Kauffman},
  \citenamefont {Landen},\ and\ \citenamefont {Suter}}]{Lindl2004The}%
  \BibitemOpen
  \bibfield  {author} {\bibinfo {author} {\bibfnamefont {J.~D.}\ \bibnamefont
  {Lindl}}, \bibinfo {author} {\bibfnamefont {P.}~\bibnamefont {Amendt}},
  \bibinfo {author} {\bibfnamefont {R.~L.}\ \bibnamefont {Berger}}, \bibinfo
  {author} {\bibfnamefont {S.~G.}\ \bibnamefont {Glendinning}}, \bibinfo
  {author} {\bibfnamefont {S.~H.}\ \bibnamefont {Glenzer}}, \bibinfo {author}
  {\bibfnamefont {S.~W.}\ \bibnamefont {Haan}}, \bibinfo {author}
  {\bibfnamefont {R.~L.}\ \bibnamefont {Kauffman}}, \bibinfo {author}
  {\bibfnamefont {O.~L.}\ \bibnamefont {Landen}}, \ and\ \bibinfo {author}
  {\bibfnamefont {L.~J.}\ \bibnamefont {Suter}},\ }\href@noop {} {\bibfield
  {journal} {\bibinfo  {journal} {Physics of Plasmas}\ }\textbf {\bibinfo
  {volume} {11}},\ \bibinfo {pages} {339} (\bibinfo {year} {2004})}\BibitemShut
  {NoStop}%
\bibitem [{\citenamefont {Michel}\ \emph {et~al.}(2009)\citenamefont {Michel},
  \citenamefont {Divol}, \citenamefont {Williams}, \citenamefont {Weber},
  \citenamefont {Thomas}, \citenamefont {Callahan}, \citenamefont {Haan},
  \citenamefont {Salmonson}, \citenamefont {Dixit},\ and\ \citenamefont
  {Hinkel}}]{Michel2009Tuning}%
  \BibitemOpen
  \bibfield  {author} {\bibinfo {author} {\bibfnamefont {P.}~\bibnamefont
  {Michel}}, \bibinfo {author} {\bibfnamefont {L.}~\bibnamefont {Divol}},
  \bibinfo {author} {\bibfnamefont {E.~A.}\ \bibnamefont {Williams}}, \bibinfo
  {author} {\bibfnamefont {S.}~\bibnamefont {Weber}}, \bibinfo {author}
  {\bibfnamefont {C.~A.}\ \bibnamefont {Thomas}}, \bibinfo {author}
  {\bibfnamefont {D.~A.}\ \bibnamefont {Callahan}}, \bibinfo {author}
  {\bibfnamefont {S.~W.}\ \bibnamefont {Haan}}, \bibinfo {author}
  {\bibfnamefont {J.~D.}\ \bibnamefont {Salmonson}}, \bibinfo {author}
  {\bibfnamefont {S.}~\bibnamefont {Dixit}}, \ and\ \bibinfo {author}
  {\bibfnamefont {D.~E.}\ \bibnamefont {Hinkel}},\ }\href@noop {} {\bibfield
  {journal} {\bibinfo  {journal} {Physical Review Letters}\ }\textbf {\bibinfo
  {volume} {102}},\ \bibinfo {pages} {025004} (\bibinfo {year}
  {2009})}\BibitemShut {NoStop}%
\bibitem [{\citenamefont {Moody}\ \emph {et~al.}(2012)\citenamefont {Moody},
  \citenamefont {Michel}, \citenamefont {Divol}, \citenamefont {Berger},
  \citenamefont {Bond}, \citenamefont {Bradley}, \citenamefont {Callahan},
  \citenamefont {Dewald}, \citenamefont {Dixit},\ and\ \citenamefont
  {Edwards}}]{Moody2012Multistep}%
  \BibitemOpen
  \bibfield  {author} {\bibinfo {author} {\bibfnamefont {J.~D.}\ \bibnamefont
  {Moody}}, \bibinfo {author} {\bibfnamefont {P.}~\bibnamefont {Michel}},
  \bibinfo {author} {\bibfnamefont {L.}~\bibnamefont {Divol}}, \bibinfo
  {author} {\bibfnamefont {R.~L.}\ \bibnamefont {Berger}}, \bibinfo {author}
  {\bibfnamefont {E.}~\bibnamefont {Bond}}, \bibinfo {author} {\bibfnamefont
  {D.~K.}\ \bibnamefont {Bradley}}, \bibinfo {author} {\bibfnamefont {D.~A.}\
  \bibnamefont {Callahan}}, \bibinfo {author} {\bibfnamefont {E.~L.}\
  \bibnamefont {Dewald}}, \bibinfo {author} {\bibfnamefont {S.}~\bibnamefont
  {Dixit}}, \ and\ \bibinfo {author} {\bibfnamefont {M.~J.}\ \bibnamefont
  {Edwards}},\ }\href@noop {} {\bibfield  {journal} {\bibinfo  {journal}
  {Nature Physics}\ }\textbf {\bibinfo {volume} {8}},\ \bibinfo {pages} {344}
  (\bibinfo {year} {2012})}\BibitemShut {NoStop}%
\bibitem [{\citenamefont {Skupsky}\ \emph {et~al.}(1989)\citenamefont
  {Skupsky}, \citenamefont {Short}, \citenamefont {Kessler}, \citenamefont
  {Craxton}, \citenamefont {Letzring},\ and\ \citenamefont
  {Soures}}]{Skupsky1989Improved}%
  \BibitemOpen
  \bibfield  {author} {\bibinfo {author} {\bibfnamefont {S.}~\bibnamefont
  {Skupsky}}, \bibinfo {author} {\bibfnamefont {R.~W.}\ \bibnamefont {Short}},
  \bibinfo {author} {\bibfnamefont {T.}~\bibnamefont {Kessler}}, \bibinfo
  {author} {\bibfnamefont {R.~S.}\ \bibnamefont {Craxton}}, \bibinfo {author}
  {\bibfnamefont {S.}~\bibnamefont {Letzring}}, \ and\ \bibinfo {author}
  {\bibfnamefont {J.~M.}\ \bibnamefont {Soures}},\ }\href@noop {} {\bibfield
  {journal} {\bibinfo  {journal} {Journal of Applied Physics}\ }\textbf
  {\bibinfo {volume} {66}},\ \bibinfo {pages} {3456} (\bibinfo {year}
  {1989})}\BibitemShut {NoStop}%
\bibitem [{\citenamefont {Dorrer}(2014)}]{Dorrer:14}%
  \BibitemOpen
  \bibfield  {author} {\bibinfo {author} {\bibfnamefont {C.}~\bibnamefont
  {Dorrer}},\ }\href {\doibase 10.1364/AO.53.001007} {\bibfield  {journal}
  {\bibinfo  {journal} {Appl. Opt.}\ }\textbf {\bibinfo {volume} {53}},\
  \bibinfo {pages} {1007} (\bibinfo {year} {2014})}\BibitemShut {NoStop}%
\bibitem [{\citenamefont {Zhong}\ \emph {et~al.}(2018)\citenamefont {Zhong},
  \citenamefont {Yi}, \citenamefont {Sui}, \citenamefont {Zhang}, \citenamefont
  {Zhang},\ and\ \citenamefont {Yuan}}]{Zhong:18}%
  \BibitemOpen
  \bibfield  {author} {\bibinfo {author} {\bibfnamefont {Z.}~\bibnamefont
  {Zhong}}, \bibinfo {author} {\bibfnamefont {M.}~\bibnamefont {Yi}}, \bibinfo
  {author} {\bibfnamefont {Z.}~\bibnamefont {Sui}}, \bibinfo {author}
  {\bibfnamefont {X.}~\bibnamefont {Zhang}}, \bibinfo {author} {\bibfnamefont
  {B.}~\bibnamefont {Zhang}}, \ and\ \bibinfo {author} {\bibfnamefont
  {X.}~\bibnamefont {Yuan}},\ }\href@noop {} {\bibfield  {journal} {\bibinfo
  {journal} {Opt. Lett.}\ }\textbf {\bibinfo {volume} {43}},\ \bibinfo {pages}
  {3285} (\bibinfo {year} {2018})}\BibitemShut {NoStop}%
\bibitem [{\citenamefont {Rothenberg}(2000)}]{Rothenberg2000Polarization}%
  \BibitemOpen
  \bibfield  {author} {\bibinfo {author} {\bibfnamefont {J.~E.}\ \bibnamefont
  {Rothenberg}},\ }\href@noop {} {\bibfield  {journal} {\bibinfo  {journal}
  {Journal of Applied Physics}\ }\textbf {\bibinfo {volume} {87}},\ \bibinfo
  {pages} {3654} (\bibinfo {year} {2000})}\BibitemShut {NoStop}%
\bibitem [{\citenamefont {Kato}\ \emph {et~al.}(1984)\citenamefont {Kato},
  \citenamefont {Mima}, \citenamefont {Miyanaga}, \citenamefont {Arinaga},
  \citenamefont {Kitagawa}, \citenamefont {Nakatsuka},\ and\ \citenamefont
  {Yamanaka}}]{PhysRevLett.53.1057}%
  \BibitemOpen
  \bibfield  {author} {\bibinfo {author} {\bibfnamefont {Y.}~\bibnamefont
  {Kato}}, \bibinfo {author} {\bibfnamefont {K.}~\bibnamefont {Mima}}, \bibinfo
  {author} {\bibfnamefont {N.}~\bibnamefont {Miyanaga}}, \bibinfo {author}
  {\bibfnamefont {S.}~\bibnamefont {Arinaga}}, \bibinfo {author} {\bibfnamefont
  {Y.}~\bibnamefont {Kitagawa}}, \bibinfo {author} {\bibfnamefont
  {M.}~\bibnamefont {Nakatsuka}}, \ and\ \bibinfo {author} {\bibfnamefont
  {C.}~\bibnamefont {Yamanaka}},\ }\href {\doibase 10.1103/PhysRevLett.53.1057}
  {\bibfield  {journal} {\bibinfo  {journal} {Phys. Rev. Lett.}\ }\textbf
  {\bibinfo {volume} {53}},\ \bibinfo {pages} {1057} (\bibinfo {year}
  {1984})}\BibitemShut {NoStop}%
\bibitem [{\citenamefont {Lin}\ \emph {et~al.}(1995)\citenamefont {Lin},
  \citenamefont {Kessler},\ and\ \citenamefont {Lawrence}}]{Lin:95}%
  \BibitemOpen
  \bibfield  {author} {\bibinfo {author} {\bibfnamefont {Y.}~\bibnamefont
  {Lin}}, \bibinfo {author} {\bibfnamefont {T.~J.}\ \bibnamefont {Kessler}}, \
  and\ \bibinfo {author} {\bibfnamefont {G.~N.}\ \bibnamefont {Lawrence}},\
  }\href {\doibase 10.1364/OL.20.000764} {\bibfield  {journal} {\bibinfo
  {journal} {Opt. Lett.}\ }\textbf {\bibinfo {volume} {20}},\ \bibinfo {pages}
  {764} (\bibinfo {year} {1995})}\BibitemShut {NoStop}%
\bibitem [{\citenamefont {Obenschain}\ \emph {et~al.}(1986)\citenamefont
  {Obenschain}, \citenamefont {Grun}, \citenamefont {Herbst}, \citenamefont
  {Kearney}, \citenamefont {Manka}, \citenamefont {McLean}, \citenamefont
  {Mostovych}, \citenamefont {Stamper}, \citenamefont {Whitlock}, \citenamefont
  {Bodner}, \citenamefont {Gardner},\ and\ \citenamefont
  {Lehmberg}}]{PhysRevLett.56.2807}%
  \BibitemOpen
  \bibfield  {author} {\bibinfo {author} {\bibfnamefont {S.~P.}\ \bibnamefont
  {Obenschain}}, \bibinfo {author} {\bibfnamefont {J.}~\bibnamefont {Grun}},
  \bibinfo {author} {\bibfnamefont {M.~J.}\ \bibnamefont {Herbst}}, \bibinfo
  {author} {\bibfnamefont {K.~J.}\ \bibnamefont {Kearney}}, \bibinfo {author}
  {\bibfnamefont {C.~K.}\ \bibnamefont {Manka}}, \bibinfo {author}
  {\bibfnamefont {E.~A.}\ \bibnamefont {McLean}}, \bibinfo {author}
  {\bibfnamefont {A.~N.}\ \bibnamefont {Mostovych}}, \bibinfo {author}
  {\bibfnamefont {J.~A.}\ \bibnamefont {Stamper}}, \bibinfo {author}
  {\bibfnamefont {R.~R.}\ \bibnamefont {Whitlock}}, \bibinfo {author}
  {\bibfnamefont {S.~E.}\ \bibnamefont {Bodner}}, \bibinfo {author}
  {\bibfnamefont {J.~H.}\ \bibnamefont {Gardner}}, \ and\ \bibinfo {author}
  {\bibfnamefont {R.~H.}\ \bibnamefont {Lehmberg}},\ }\href {\doibase
  10.1103/PhysRevLett.56.2807} {\bibfield  {journal} {\bibinfo  {journal}
  {Phys. Rev. Lett.}\ }\textbf {\bibinfo {volume} {56}},\ \bibinfo {pages}
  {2807} (\bibinfo {year} {1986})}\BibitemShut {NoStop}%
\bibitem [{\citenamefont {Deng}\ \emph {et~al.}(1986)\citenamefont {Deng},
  \citenamefont {Liang}, \citenamefont {Chen}, \citenamefont {Yu},\ and\
  \citenamefont {Ma}}]{Deng:86}%
  \BibitemOpen
  \bibfield  {author} {\bibinfo {author} {\bibfnamefont {X.}~\bibnamefont
  {Deng}}, \bibinfo {author} {\bibfnamefont {X.}~\bibnamefont {Liang}},
  \bibinfo {author} {\bibfnamefont {Z.}~\bibnamefont {Chen}}, \bibinfo {author}
  {\bibfnamefont {W.}~\bibnamefont {Yu}}, \ and\ \bibinfo {author}
  {\bibfnamefont {R.}~\bibnamefont {Ma}},\ }\href {\doibase
  10.1364/AO.25.000377} {\bibfield  {journal} {\bibinfo  {journal} {Appl.
  Opt.}\ }\textbf {\bibinfo {volume} {25}},\ \bibinfo {pages} {377} (\bibinfo
  {year} {1986})}\BibitemShut {NoStop}%
\bibitem [{\citenamefont {Spaeth}\ \emph {et~al.}(2016)\citenamefont {Spaeth},
  \citenamefont {Manes}, \citenamefont {Bowers}, \citenamefont {Celliers},
  \citenamefont {Nicola}, \citenamefont {Nicola}, \citenamefont {Dixit},
  \citenamefont {Erbert}, \citenamefont {Heebner},\ and\ \citenamefont
  {Kalantar}}]{Spaeth2016National}%
  \BibitemOpen
  \bibfield  {author} {\bibinfo {author} {\bibfnamefont {M.~L.}\ \bibnamefont
  {Spaeth}}, \bibinfo {author} {\bibfnamefont {K.~R.}\ \bibnamefont {Manes}},
  \bibinfo {author} {\bibfnamefont {M.}~\bibnamefont {Bowers}}, \bibinfo
  {author} {\bibfnamefont {P.}~\bibnamefont {Celliers}}, \bibinfo {author}
  {\bibfnamefont {J.-M.~D.}\ \bibnamefont {Nicola}}, \bibinfo {author}
  {\bibfnamefont {P.~D.}\ \bibnamefont {Nicola}}, \bibinfo {author}
  {\bibfnamefont {S.}~\bibnamefont {Dixit}}, \bibinfo {author} {\bibfnamefont
  {G.}~\bibnamefont {Erbert}}, \bibinfo {author} {\bibfnamefont
  {J.}~\bibnamefont {Heebner}}, \ and\ \bibinfo {author} {\bibfnamefont
  {D.}~\bibnamefont {Kalantar}},\ }\href@noop {} {\bibfield  {journal}
  {\bibinfo  {journal} {Fusion Science \& Technology}\ }\textbf {\bibinfo
  {volume} {69}},\ \bibinfo {pages} {366} (\bibinfo {year} {2016})}\BibitemShut
  {NoStop}%
\bibitem [{\citenamefont {Dorrer}\ \emph {et~al.}(2017)\citenamefont {Dorrer},
  \citenamefont {Consentino}, \citenamefont {Cuffney}, \citenamefont
  {Begishev}, \citenamefont {Hill},\ and\ \citenamefont
  {Bromage}}]{Dorrer2017Spectrally}%
  \BibitemOpen
  \bibfield  {author} {\bibinfo {author} {\bibfnamefont {C.}~\bibnamefont
  {Dorrer}}, \bibinfo {author} {\bibfnamefont {A.}~\bibnamefont {Consentino}},
  \bibinfo {author} {\bibfnamefont {R.}~\bibnamefont {Cuffney}}, \bibinfo
  {author} {\bibfnamefont {I.~A.}\ \bibnamefont {Begishev}}, \bibinfo {author}
  {\bibfnamefont {E.~M.}\ \bibnamefont {Hill}}, \ and\ \bibinfo {author}
  {\bibfnamefont {J.}~\bibnamefont {Bromage}},\ }\href@noop {} {\bibfield
  {journal} {\bibinfo  {journal} {Optics Express}\ }\textbf {\bibinfo {volume}
  {25}},\ \bibinfo {pages} {26802} (\bibinfo {year} {2017})}\BibitemShut
  {NoStop}%
\bibitem [{\citenamefont {Fleurot}\ \emph {et~al.}(2005)\citenamefont
  {Fleurot}, \citenamefont {Cavailler},\ and\ \citenamefont
  {Bourgade}}]{FLEUROT2005147}%
  \BibitemOpen
  \bibfield  {author} {\bibinfo {author} {\bibfnamefont {N.}~\bibnamefont
  {Fleurot}}, \bibinfo {author} {\bibfnamefont {C.}~\bibnamefont {Cavailler}},
  \ and\ \bibinfo {author} {\bibfnamefont {J.}~\bibnamefont {Bourgade}},\
  }\href@noop {} {\bibfield  {journal} {\bibinfo  {journal} {Fusion Engineering
  and Design}\ }\textbf {\bibinfo {volume} {74}},\ \bibinfo {pages} {147 }
  (\bibinfo {year} {2005})}\BibitemShut {NoStop}%
\bibitem [{\citenamefont {Jiang}\ \emph {et~al.}(2018)\citenamefont {Jiang},
  \citenamefont {Wang}, \citenamefont {Ding}, \citenamefont {Liu},
  \citenamefont {Yang}, \citenamefont {Li}, \citenamefont {Huang},
  \citenamefont {Cao}, \citenamefont {Yang}, \citenamefont {Hu}, \citenamefont
  {Miao}, \citenamefont {Zhang}, \citenamefont {Wang}, \citenamefont {Yang},
  \citenamefont {Rongqing}, \citenamefont {Tang}, \citenamefont {Kuang},
  \citenamefont {Li}, \citenamefont {Dong},\ and\ \citenamefont
  {Zhang}}]{Jiang2018Experimental}%
  \BibitemOpen
  \bibfield  {author} {\bibinfo {author} {\bibfnamefont {S.}~\bibnamefont
  {Jiang}}, \bibinfo {author} {\bibfnamefont {F.}~\bibnamefont {Wang}},
  \bibinfo {author} {\bibfnamefont {Y.}~\bibnamefont {Ding}}, \bibinfo {author}
  {\bibfnamefont {S.}~\bibnamefont {Liu}}, \bibinfo {author} {\bibfnamefont
  {J.}~\bibnamefont {Yang}}, \bibinfo {author} {\bibfnamefont {S.}~\bibnamefont
  {Li}}, \bibinfo {author} {\bibfnamefont {T.}~\bibnamefont {Huang}}, \bibinfo
  {author} {\bibfnamefont {Z.}~\bibnamefont {Cao}}, \bibinfo {author}
  {\bibfnamefont {Z.}~\bibnamefont {Yang}}, \bibinfo {author} {\bibfnamefont
  {X.}~\bibnamefont {Hu}}, \bibinfo {author} {\bibfnamefont {W.}~\bibnamefont
  {Miao}}, \bibinfo {author} {\bibfnamefont {J.}~\bibnamefont {Zhang}},
  \bibinfo {author} {\bibfnamefont {Z.}~\bibnamefont {Wang}}, \bibinfo {author}
  {\bibfnamefont {G.}~\bibnamefont {Yang}}, \bibinfo {author} {\bibfnamefont
  {Y.}~\bibnamefont {Rongqing}}, \bibinfo {author} {\bibfnamefont
  {Q.}~\bibnamefont {Tang}}, \bibinfo {author} {\bibfnamefont {L.}~\bibnamefont
  {Kuang}}, \bibinfo {author} {\bibfnamefont {Z.}~\bibnamefont {Li}}, \bibinfo
  {author} {\bibfnamefont {Y.}~\bibnamefont {Dong}}, \ and\ \bibinfo {author}
  {\bibfnamefont {B.}~\bibnamefont {Zhang}},\ }\href@noop {} {\bibfield
  {journal} {\bibinfo  {journal} {Nuclear Fusion}\ }\textbf {\bibinfo {volume}
  {59}} (\bibinfo {year} {2018})}\BibitemShut {NoStop}%
  \bibitem [{\citenamefont {Haynam}\ \emph {et~al.}(2007)\citenamefont {Haynam},
  \citenamefont {Wegner}, \citenamefont {Auerbach}, \citenamefont {Bowers},
  \citenamefont {Dixit}, \citenamefont {Erbert}, \citenamefont {Heestand},
  \citenamefont {Henesian}, \citenamefont {Hermann}, \citenamefont {Jancaitis},
  \citenamefont {Manes}, \citenamefont {Marshall}, \citenamefont {Mehta},
  \citenamefont {Menapace}, \citenamefont {Moses}, \citenamefont {Murray},
  \citenamefont {Nostrand}, \citenamefont {Orth}, \citenamefont {Patterson},
  \citenamefont {Sacks}, \citenamefont {Shaw}, \citenamefont {Spaeth},
  \citenamefont {Sutton}, \citenamefont {Williams}, \citenamefont {Widmayer},
  \citenamefont {White}, \citenamefont {Yang},\ and\ \citenamefont
  {Wonterghem}}]{Haynam:07}%
  \BibitemOpen
  \bibfield  {author} {\bibinfo {author} {\bibfnamefont {C.~A.}\ \bibnamefont
  {Haynam}}, \bibinfo {author} {\bibfnamefont {P.~J.}\ \bibnamefont {Wegner}},
  \bibinfo {author} {\bibfnamefont {J.~M.}\ \bibnamefont {Auerbach}}, \bibinfo
  {author} {\bibfnamefont {M.~W.}\ \bibnamefont {Bowers}}, \bibinfo {author}
  {\bibfnamefont {S.~N.}\ \bibnamefont {Dixit}}, \bibinfo {author}
  {\bibfnamefont {G.~V.}\ \bibnamefont {Erbert}}, \bibinfo {author}
  {\bibfnamefont {G.~M.}\ \bibnamefont {Heestand}}, \bibinfo {author}
  {\bibfnamefont {M.~A.}\ \bibnamefont {Henesian}}, \bibinfo {author}
  {\bibfnamefont {M.~R.}\ \bibnamefont {Hermann}}, \bibinfo {author}
  {\bibfnamefont {K.~S.}\ \bibnamefont {Jancaitis}}, \bibinfo {author}
  {\bibfnamefont {K.~R.}\ \bibnamefont {Manes}}, \bibinfo {author}
  {\bibfnamefont {C.~D.}\ \bibnamefont {Marshall}}, \bibinfo {author}
  {\bibfnamefont {N.~C.}\ \bibnamefont {Mehta}}, \bibinfo {author}
  {\bibfnamefont {J.}~\bibnamefont {Menapace}}, \bibinfo {author}
  {\bibfnamefont {E.}~\bibnamefont {Moses}}, \bibinfo {author} {\bibfnamefont
  {J.~R.}\ \bibnamefont {Murray}}, \bibinfo {author} {\bibfnamefont {M.~C.}\
  \bibnamefont {Nostrand}}, \bibinfo {author} {\bibfnamefont {C.~D.}\
  \bibnamefont {Orth}}, \bibinfo {author} {\bibfnamefont {R.}~\bibnamefont
  {Patterson}}, \bibinfo {author} {\bibfnamefont {R.~A.}\ \bibnamefont
  {Sacks}}, \bibinfo {author} {\bibfnamefont {M.~J.}\ \bibnamefont {Shaw}},
  \bibinfo {author} {\bibfnamefont {M.}~\bibnamefont {Spaeth}}, \bibinfo
  {author} {\bibfnamefont {S.~B.}\ \bibnamefont {Sutton}}, \bibinfo {author}
  {\bibfnamefont {W.~H.}\ \bibnamefont {Williams}}, \bibinfo {author}
  {\bibfnamefont {C.~C.}\ \bibnamefont {Widmayer}}, \bibinfo {author}
  {\bibfnamefont {R.~K.}\ \bibnamefont {White}}, \bibinfo {author}
  {\bibfnamefont {S.~T.}\ \bibnamefont {Yang}}, \ and\ \bibinfo {author}
  {\bibfnamefont {B.~M.~V.}\ \bibnamefont {Wonterghem}},\ }\href {\doibase
  10.1364/AO.46.003276} {\bibfield  {journal} {\bibinfo  {journal} {Appl.
  Opt.}\ }\textbf {\bibinfo {volume} {46}},\ \bibinfo {pages} {3276} (\bibinfo
  {year} {2007})}\BibitemShut {NoStop}%
\bibitem [{\citenamefont {Montgomery}(2016)}]{doi:10.1063/1.4946016}%
  \BibitemOpen
  \bibfield  {author} {\bibinfo {author} {\bibfnamefont {D.~S.}\ \bibnamefont
  {Montgomery}},\ }\href@noop {} {\bibfield  {journal} {\bibinfo  {journal}
  {Physics of Plasmas}\ }\textbf {\bibinfo {volume} {23}},\ \bibinfo {pages}
  {055601} (\bibinfo {year} {2016})}\BibitemShut {NoStop}%
\bibitem [{\citenamefont {Rosenberg}\ \emph {et~al.}(2018)\citenamefont
  {Rosenberg}, \citenamefont {Solodov}, \citenamefont {Myatt}, \citenamefont
  {Seka}, \citenamefont {Michel}, \citenamefont {Hohenberger}, \citenamefont
  {Short}, \citenamefont {Epstein}, \citenamefont {Regan}, \citenamefont
  {Campbell}, \citenamefont {Chapman}, \citenamefont {Goyon}, \citenamefont
  {Ralph}, \citenamefont {Barrios}, \citenamefont {Moody},\ and\ \citenamefont
  {Bates}}]{PhysRevLett.120.055001}%
  \BibitemOpen
  \bibfield  {author} {\bibinfo {author} {\bibfnamefont {M.~J.}\ \bibnamefont
  {Rosenberg}}, \bibinfo {author} {\bibfnamefont {A.~A.}\ \bibnamefont
  {Solodov}}, \bibinfo {author} {\bibfnamefont {J.~F.}\ \bibnamefont {Myatt}},
  \bibinfo {author} {\bibfnamefont {W.}~\bibnamefont {Seka}}, \bibinfo {author}
  {\bibfnamefont {P.}~\bibnamefont {Michel}}, \bibinfo {author} {\bibfnamefont
  {M.}~\bibnamefont {Hohenberger}}, \bibinfo {author} {\bibfnamefont {R.~W.}\
  \bibnamefont {Short}}, \bibinfo {author} {\bibfnamefont {R.}~\bibnamefont
  {Epstein}}, \bibinfo {author} {\bibfnamefont {S.~P.}\ \bibnamefont {Regan}},
  \bibinfo {author} {\bibfnamefont {E.~M.}\ \bibnamefont {Campbell}}, \bibinfo
  {author} {\bibfnamefont {T.}~\bibnamefont {Chapman}}, \bibinfo {author}
  {\bibfnamefont {C.}~\bibnamefont {Goyon}}, \bibinfo {author} {\bibfnamefont
  {J.~E.}\ \bibnamefont {Ralph}}, \bibinfo {author} {\bibfnamefont {M.~A.}\
  \bibnamefont {Barrios}}, \bibinfo {author} {\bibfnamefont {J.~D.}\
  \bibnamefont {Moody}}, \ and\ \bibinfo {author} {\bibfnamefont {J.~W.}\
  \bibnamefont {Bates}},\ }\href@noop {} {\bibfield  {journal} {\bibinfo
  {journal} {Phys. Rev. Lett.}\ }\textbf {\bibinfo {volume} {120}},\ \bibinfo
  {pages} {055001} (\bibinfo {year} {2018})}\BibitemShut {NoStop}%
\bibitem [{\citenamefont {Michel}\ \emph {et~al.}(2015)\citenamefont {Michel},
  \citenamefont {Divol}, \citenamefont {Dewald}, \citenamefont {Milovich},
  \citenamefont {Hohenberger}, \citenamefont {Jones}, \citenamefont {Hopkins},
  \citenamefont {Berger}, \citenamefont {Kruer},\ and\ \citenamefont
  {Moody}}]{PhysRevLett.115.055003}%
  \BibitemOpen
  \bibfield  {author} {\bibinfo {author} {\bibfnamefont {P.}~\bibnamefont
  {Michel}}, \bibinfo {author} {\bibfnamefont {L.}~\bibnamefont {Divol}},
  \bibinfo {author} {\bibfnamefont {E.~L.}\ \bibnamefont {Dewald}}, \bibinfo
  {author} {\bibfnamefont {J.~L.}\ \bibnamefont {Milovich}}, \bibinfo {author}
  {\bibfnamefont {M.}~\bibnamefont {Hohenberger}}, \bibinfo {author}
  {\bibfnamefont {O.~S.}\ \bibnamefont {Jones}}, \bibinfo {author}
  {\bibfnamefont {L.~B.}\ \bibnamefont {Hopkins}}, \bibinfo {author}
  {\bibfnamefont {R.~L.}\ \bibnamefont {Berger}}, \bibinfo {author}
  {\bibfnamefont {W.~L.}\ \bibnamefont {Kruer}}, \ and\ \bibinfo {author}
  {\bibfnamefont {J.~D.}\ \bibnamefont {Moody}},\ }\href {\doibase
  10.1103/PhysRevLett.115.055003} {\bibfield  {journal} {\bibinfo  {journal}
  {Phys. Rev. Lett.}\ }\textbf {\bibinfo {volume} {115}},\ \bibinfo {pages}
  {055003} (\bibinfo {year} {2015})}\BibitemShut {NoStop}%
\bibitem [{\citenamefont {Eimerl}\ \emph {et~al.}(2014)\citenamefont {Eimerl},
  \citenamefont {Campbell}, \citenamefont {Krupke}, \citenamefont {Zweiback},
  \citenamefont {Kruer}, \citenamefont {Marozas}, \citenamefont {Zuegel},
  \citenamefont {Myatt}, \citenamefont {Kelly},\ and\ \citenamefont
  {Froula}}]{Eimerl2014StarDriver}%
  \BibitemOpen
  \bibfield  {author} {\bibinfo {author} {\bibfnamefont {D.}~\bibnamefont
  {Eimerl}}, \bibinfo {author} {\bibfnamefont {E.~M.}\ \bibnamefont
  {Campbell}}, \bibinfo {author} {\bibfnamefont {W.~F.}\ \bibnamefont
  {Krupke}}, \bibinfo {author} {\bibfnamefont {J.}~\bibnamefont {Zweiback}},
  \bibinfo {author} {\bibfnamefont {W.~L.}\ \bibnamefont {Kruer}}, \bibinfo
  {author} {\bibfnamefont {J.}~\bibnamefont {Marozas}}, \bibinfo {author}
  {\bibfnamefont {J.}~\bibnamefont {Zuegel}}, \bibinfo {author} {\bibfnamefont
  {J.}~\bibnamefont {Myatt}}, \bibinfo {author} {\bibfnamefont
  {J.}~\bibnamefont {Kelly}}, \ and\ \bibinfo {author} {\bibfnamefont
  {D.}~\bibnamefont {Froula}},\ }\href@noop {} {\bibfield  {journal} {\bibinfo
  {journal} {Journal of Fusion Energy}\ }\textbf {\bibinfo {volume} {33}},\
  \bibinfo {pages} {476} (\bibinfo {year} {2014})}\BibitemShut {NoStop}%
\bibitem [{\citenamefont {Eimerl}\ \emph {et~al.}(1993)\citenamefont {Eimerl},
  \citenamefont {Milam},\ and\ \citenamefont {Yu}}]{PhysRevLett.70.2738}%
  \BibitemOpen
  \bibfield  {author} {\bibinfo {author} {\bibfnamefont {D.}~\bibnamefont
  {Eimerl}}, \bibinfo {author} {\bibfnamefont {D.}~\bibnamefont {Milam}}, \
  and\ \bibinfo {author} {\bibfnamefont {J.}~\bibnamefont {Yu}},\ }\href
  {\doibase 10.1103/PhysRevLett.70.2738} {\bibfield  {journal} {\bibinfo
  {journal} {Phys. Rev. Lett.}\ }\textbf {\bibinfo {volume} {70}},\ \bibinfo
  {pages} {2738} (\bibinfo {year} {1993})}\BibitemShut {NoStop}%
\bibitem [{\citenamefont {Bates}\ \emph {et~al.}(2018)\citenamefont {Bates},
  \citenamefont {Myatt}, \citenamefont {Shaw}, \citenamefont {Follett},
  \citenamefont {Weaver}, \citenamefont {Lehmberg},\ and\ \citenamefont
  {Obenschain}}]{PhysRevE.97.061202}%
  \BibitemOpen
  \bibfield  {author} {\bibinfo {author} {\bibfnamefont {J.~W.}\ \bibnamefont
  {Bates}}, \bibinfo {author} {\bibfnamefont {J.~F.}\ \bibnamefont {Myatt}},
  \bibinfo {author} {\bibfnamefont {J.~G.}\ \bibnamefont {Shaw}}, \bibinfo
  {author} {\bibfnamefont {R.~K.}\ \bibnamefont {Follett}}, \bibinfo {author}
  {\bibfnamefont {J.~L.}\ \bibnamefont {Weaver}}, \bibinfo {author}
  {\bibfnamefont {R.~H.}\ \bibnamefont {Lehmberg}}, \ and\ \bibinfo {author}
  {\bibfnamefont {S.~P.}\ \bibnamefont {Obenschain}},\ }\href@noop {}
  {\bibfield  {journal} {\bibinfo  {journal} {Phys. Rev. E}\ }\textbf {\bibinfo
  {volume} {97}},\ \bibinfo {pages} {061202} (\bibinfo {year}
  {2018})}\BibitemShut {NoStop}%
\bibitem [{\citenamefont {Manes}\ \emph {et~al.}(2016)\citenamefont {Manes},
  \citenamefont {Spaeth}, \citenamefont {Adams}, \citenamefont {Bowers},
  \citenamefont {Bude}, \citenamefont {Carr}, \citenamefont {Conder},
  \citenamefont {Cross}, \citenamefont {Demos}, \citenamefont {{Di Nicola}},
  \citenamefont {Dixit}, \citenamefont {Feigenbaum}, \citenamefont {Finucane},
  \citenamefont {Guss}, \citenamefont {Henesian}, \citenamefont {Honig},
  \citenamefont {Kalantar}, \citenamefont {Kegelmeyer}, \citenamefont {Liao},
  \citenamefont {MacGowan}, \citenamefont {Matthews}, \citenamefont
  {McCandless}, \citenamefont {Mehta}, \citenamefont {Miller}, \citenamefont
  {Negres}, \citenamefont {Norton}, \citenamefont {Nostrand}, \citenamefont
  {Orth}, \citenamefont {Sacks}, \citenamefont {Shaw}, \citenamefont {Siegel},
  \citenamefont {Stolz}, \citenamefont {Suratwala}, \citenamefont {Trenholme},
  \citenamefont {Wegner}, \citenamefont {Whitman}, \citenamefont {Widmayer},\
  and\ \citenamefont {Yang}}]{031b69f3114540a4811dc0cee2ed0b87}%
  \BibitemOpen
  \bibfield  {author} {\bibinfo {author} {\bibfnamefont {K.}~\bibnamefont
  {Manes}}, \bibinfo {author} {\bibfnamefont {M.}~\bibnamefont {Spaeth}},
  \bibinfo {author} {\bibfnamefont {J.}~\bibnamefont {Adams}}, \bibinfo
  {author} {\bibfnamefont {M.}~\bibnamefont {Bowers}}, \bibinfo {author}
  {\bibfnamefont {J.}~\bibnamefont {Bude}}, \bibinfo {author} {\bibfnamefont
  {C.}~\bibnamefont {Carr}}, \bibinfo {author} {\bibfnamefont {A.}~\bibnamefont
  {Conder}}, \bibinfo {author} {\bibfnamefont {D.}~\bibnamefont {Cross}},
  \bibinfo {author} {\bibfnamefont {S.}~\bibnamefont {Demos}}, \bibinfo
  {author} {\bibfnamefont {J.}~\bibnamefont {{Di Nicola}}}, \bibinfo {author}
  {\bibfnamefont {S.}~\bibnamefont {Dixit}}, \bibinfo {author} {\bibfnamefont
  {E.}~\bibnamefont {Feigenbaum}}, \bibinfo {author} {\bibfnamefont
  {R.}~\bibnamefont {Finucane}}, \bibinfo {author} {\bibfnamefont
  {G.}~\bibnamefont {Guss}}, \bibinfo {author} {\bibfnamefont {M.}~\bibnamefont
  {Henesian}}, \bibinfo {author} {\bibfnamefont {J.}~\bibnamefont {Honig}},
  \bibinfo {author} {\bibfnamefont {D.}~\bibnamefont {Kalantar}}, \bibinfo
  {author} {\bibfnamefont {L.}~\bibnamefont {Kegelmeyer}}, \bibinfo {author}
  {\bibfnamefont {Z.}~\bibnamefont {Liao}}, \bibinfo {author} {\bibfnamefont
  {B.}~\bibnamefont {MacGowan}}, \bibinfo {author} {\bibfnamefont
  {M.}~\bibnamefont {Matthews}}, \bibinfo {author} {\bibfnamefont
  {K.}~\bibnamefont {McCandless}}, \bibinfo {author} {\bibfnamefont
  {N.}~\bibnamefont {Mehta}}, \bibinfo {author} {\bibfnamefont
  {P.}~\bibnamefont {Miller}}, \bibinfo {author} {\bibfnamefont
  {R.}~\bibnamefont {Negres}}, \bibinfo {author} {\bibfnamefont
  {M.}~\bibnamefont {Norton}}, \bibinfo {author} {\bibfnamefont
  {M.}~\bibnamefont {Nostrand}}, \bibinfo {author} {\bibfnamefont
  {C.}~\bibnamefont {Orth}}, \bibinfo {author} {\bibfnamefont {R.}~\bibnamefont
  {Sacks}}, \bibinfo {author} {\bibfnamefont {M.}~\bibnamefont {Shaw}},
  \bibinfo {author} {\bibfnamefont {L.}~\bibnamefont {Siegel}}, \bibinfo
  {author} {\bibfnamefont {C.}~\bibnamefont {Stolz}}, \bibinfo {author}
  {\bibfnamefont {T.}~\bibnamefont {Suratwala}}, \bibinfo {author}
  {\bibfnamefont {J.}~\bibnamefont {Trenholme}}, \bibinfo {author}
  {\bibfnamefont {P.}~\bibnamefont {Wegner}}, \bibinfo {author} {\bibfnamefont
  {P.}~\bibnamefont {Whitman}}, \bibinfo {author} {\bibfnamefont
  {C.}~\bibnamefont {Widmayer}}, \ and\ \bibinfo {author} {\bibfnamefont
  {S.}~\bibnamefont {Yang}},\ }\href {\doibase 10.13182/FST15-139} {\bibfield
  {journal} {\bibinfo  {journal} {Fusion Science and Technology}\ }\textbf
  {\bibinfo {volume} {69}},\ \bibinfo {pages} {146} (\bibinfo {year}
  {2016})}\BibitemShut {NoStop}%
\bibitem [{\citenamefont {Chambonneau}\ \emph {et~al.}(2018)\citenamefont
  {Chambonneau}, \citenamefont {Rullier}, \citenamefont {Grua},\ and\
  \citenamefont {Lamaign\`{e}re}}]{Chambonneau:18}%
  \BibitemOpen
  \bibfield  {author} {\bibinfo {author} {\bibfnamefont {M.}~\bibnamefont
  {Chambonneau}}, \bibinfo {author} {\bibfnamefont {J.-L.}\ \bibnamefont
  {Rullier}}, \bibinfo {author} {\bibfnamefont {P.}~\bibnamefont {Grua}}, \
  and\ \bibinfo {author} {\bibfnamefont {L.}~\bibnamefont {Lamaign\`{e}re}},\
  }\href@noop {} {\bibfield  {journal} {\bibinfo  {journal} {Opt. Express}\
  }\textbf {\bibinfo {volume} {26}},\ \bibinfo {pages} {21819} (\bibinfo {year}
  {2018})}\BibitemShut {NoStop}%
\bibitem [{\citenamefont {Suter}\ \emph {et~al.}(2004)\citenamefont {Suter},
  \citenamefont {Glenzer}, \citenamefont {Haan}, \citenamefont {Hammel},
  \citenamefont {Manes}, \citenamefont {Meezan}, \citenamefont {Moody},
  \citenamefont {Spaeth}, \citenamefont {Divol}, \citenamefont {Oades},\ and\
  \citenamefont {Stevenson}}]{doi:10.1063/1.1687725}%
  \BibitemOpen
  \bibfield  {author} {\bibinfo {author} {\bibfnamefont {L.~J.}\ \bibnamefont
  {Suter}}, \bibinfo {author} {\bibfnamefont {S.}~\bibnamefont {Glenzer}},
  \bibinfo {author} {\bibfnamefont {S.}~\bibnamefont {Haan}}, \bibinfo {author}
  {\bibfnamefont {B.}~\bibnamefont {Hammel}}, \bibinfo {author} {\bibfnamefont
  {K.}~\bibnamefont {Manes}}, \bibinfo {author} {\bibfnamefont
  {N.}~\bibnamefont {Meezan}}, \bibinfo {author} {\bibfnamefont
  {J.}~\bibnamefont {Moody}}, \bibinfo {author} {\bibfnamefont
  {M.}~\bibnamefont {Spaeth}}, \bibinfo {author} {\bibfnamefont
  {L.}~\bibnamefont {Divol}}, \bibinfo {author} {\bibfnamefont
  {K.}~\bibnamefont {Oades}}, \ and\ \bibinfo {author} {\bibfnamefont
  {M.}~\bibnamefont {Stevenson}},\ }\href@noop {} {\bibfield  {journal}
  {\bibinfo  {journal} {Physics of Plasmas}\ }\textbf {\bibinfo {volume}
  {11}},\ \bibinfo {pages} {2738} (\bibinfo {year} {2004})}\BibitemShut
  {NoStop}%
\bibitem [{\citenamefont {Heestand}\ \emph {et~al.}(2008)\citenamefont
  {Heestand}, \citenamefont {Haynam}, \citenamefont {Wegner}, \citenamefont
  {Bowers}, \citenamefont {Dixit}, \citenamefont {Erbert}, \citenamefont
  {Henesian}, \citenamefont {Hermann}, \citenamefont {Jancaitis}, \citenamefont
  {Knittel}, \citenamefont {Kohut}, \citenamefont {Lindl}, \citenamefont
  {Manes}, \citenamefont {Marshall}, \citenamefont {Mehta}, \citenamefont
  {Menapace}, \citenamefont {Moses}, \citenamefont {Murray}, \citenamefont
  {Nostrand}, \citenamefont {Orth}, \citenamefont {Patterson}, \citenamefont
  {Sacks}, \citenamefont {Saunders}, \citenamefont {Shaw}, \citenamefont
  {Spaeth}, \citenamefont {Sutton}, \citenamefont {Williams}, \citenamefont
  {Widmayer}, \citenamefont {White}, \citenamefont {Whitman}, \citenamefont
  {Yang},\ and\ \citenamefont {Wonterghem}}]{Heestand:08}%
  \BibitemOpen
  \bibfield  {author} {\bibinfo {author} {\bibfnamefont {G.~M.}\ \bibnamefont
  {Heestand}}, \bibinfo {author} {\bibfnamefont {C.~A.}\ \bibnamefont
  {Haynam}}, \bibinfo {author} {\bibfnamefont {P.~J.}\ \bibnamefont {Wegner}},
  \bibinfo {author} {\bibfnamefont {M.~W.}\ \bibnamefont {Bowers}}, \bibinfo
  {author} {\bibfnamefont {S.~N.}\ \bibnamefont {Dixit}}, \bibinfo {author}
  {\bibfnamefont {G.~V.}\ \bibnamefont {Erbert}}, \bibinfo {author}
  {\bibfnamefont {M.~A.}\ \bibnamefont {Henesian}}, \bibinfo {author}
  {\bibfnamefont {M.~R.}\ \bibnamefont {Hermann}}, \bibinfo {author}
  {\bibfnamefont {K.~S.}\ \bibnamefont {Jancaitis}}, \bibinfo {author}
  {\bibfnamefont {K.}~\bibnamefont {Knittel}}, \bibinfo {author} {\bibfnamefont
  {T.}~\bibnamefont {Kohut}}, \bibinfo {author} {\bibfnamefont {J.~D.}\
  \bibnamefont {Lindl}}, \bibinfo {author} {\bibfnamefont {K.~R.}\ \bibnamefont
  {Manes}}, \bibinfo {author} {\bibfnamefont {C.~D.}\ \bibnamefont {Marshall}},
  \bibinfo {author} {\bibfnamefont {N.~C.}\ \bibnamefont {Mehta}}, \bibinfo
  {author} {\bibfnamefont {J.}~\bibnamefont {Menapace}}, \bibinfo {author}
  {\bibfnamefont {E.}~\bibnamefont {Moses}}, \bibinfo {author} {\bibfnamefont
  {J.~R.}\ \bibnamefont {Murray}}, \bibinfo {author} {\bibfnamefont {M.~C.}\
  \bibnamefont {Nostrand}}, \bibinfo {author} {\bibfnamefont {C.~D.}\
  \bibnamefont {Orth}}, \bibinfo {author} {\bibfnamefont {R.}~\bibnamefont
  {Patterson}}, \bibinfo {author} {\bibfnamefont {R.~A.}\ \bibnamefont
  {Sacks}}, \bibinfo {author} {\bibfnamefont {R.}~\bibnamefont {Saunders}},
  \bibinfo {author} {\bibfnamefont {M.~J.}\ \bibnamefont {Shaw}}, \bibinfo
  {author} {\bibfnamefont {M.}~\bibnamefont {Spaeth}}, \bibinfo {author}
  {\bibfnamefont {S.~B.}\ \bibnamefont {Sutton}}, \bibinfo {author}
  {\bibfnamefont {W.~H.}\ \bibnamefont {Williams}}, \bibinfo {author}
  {\bibfnamefont {C.~C.}\ \bibnamefont {Widmayer}}, \bibinfo {author}
  {\bibfnamefont {R.~K.}\ \bibnamefont {White}}, \bibinfo {author}
  {\bibfnamefont {P.~K.}\ \bibnamefont {Whitman}}, \bibinfo {author}
  {\bibfnamefont {S.~T.}\ \bibnamefont {Yang}}, \ and\ \bibinfo {author}
  {\bibfnamefont {B.~M.~V.}\ \bibnamefont {Wonterghem}},\ }\href
  {http://ao.osa.org/abstract.cfm?URI=ao-47-19-3494} {\bibfield  {journal}
  {\bibinfo  {journal} {Appl. Opt.}\ }\textbf {\bibinfo {volume} {47}},\
  \bibinfo {pages} {3494} (\bibinfo {year} {2008})}\BibitemShut {NoStop}%
\bibitem [{\citenamefont {Glenn}(1969)}]{1081948}%
  \BibitemOpen
  \bibfield  {author} {\bibinfo {author} {\bibfnamefont {W.}~\bibnamefont
  {Glenn}},\ }\href {\doibase 10.1109/JQE.1969.1081948} {\bibfield  {journal}
  {\bibinfo  {journal} {IEEE Journal of Quantum Electronics}\ }\textbf
  {\bibinfo {volume} {5}},\ \bibinfo {pages} {284} (\bibinfo {year}
  {1969})}\BibitemShut {NoStop}%
\bibitem [{\citenamefont {Eimerl}(1987)}]{1073521}%
  \BibitemOpen
  \bibfield  {author} {\bibinfo {author} {\bibfnamefont {D.}~\bibnamefont
  {Eimerl}},\ }\href {\doibase 10.1109/JQE.1987.1073521} {\bibfield  {journal}
  {\bibinfo  {journal} {IEEE Journal of Quantum Electronics}\ }\textbf
  {\bibinfo {volume} {23}},\ \bibinfo {pages} {1361} (\bibinfo {year}
  {1987})}\BibitemShut {NoStop}%
\bibitem [{\citenamefont {Skeldon}\ \emph {et~al.}(1992)\citenamefont
  {Skeldon}, \citenamefont {Craxton}, \citenamefont {Kessler}, \citenamefont
  {Seka}, \citenamefont {Short}, \citenamefont {Skupsky},\ and\ \citenamefont
  {Soures}}]{135282}%
  \BibitemOpen
  \bibfield  {author} {\bibinfo {author} {\bibfnamefont {M.~D.}\ \bibnamefont
  {Skeldon}}, \bibinfo {author} {\bibfnamefont {R.~S.}\ \bibnamefont
  {Craxton}}, \bibinfo {author} {\bibfnamefont {T.}~\bibnamefont {Kessler}},
  \bibinfo {author} {\bibfnamefont {W.}~\bibnamefont {Seka}}, \bibinfo {author}
  {\bibfnamefont {R.~W.}\ \bibnamefont {Short}}, \bibinfo {author}
  {\bibfnamefont {S.}~\bibnamefont {Skupsky}}, \ and\ \bibinfo {author}
  {\bibfnamefont {J.~M.}\ \bibnamefont {Soures}},\ }\href {\doibase
  10.1109/3.135282} {\bibfield  {journal} {\bibinfo  {journal} {IEEE Journal of
  Quantum Electronics}\ }\textbf {\bibinfo {volume} {28}},\ \bibinfo {pages}
  {1389} (\bibinfo {year} {1992})}\BibitemShut {NoStop}%
\bibitem [{\citenamefont {Nakatsuka}\ \emph {et~al.}(1993)\citenamefont
  {Nakatsuka}, \citenamefont {Miyanaga}, \citenamefont {Kanabe}, \citenamefont
  {Nakano},\ and\ \citenamefont {Nakai}}]{Nakatsuka1993Partially}%
  \BibitemOpen
  \bibfield  {author} {\bibinfo {author} {\bibfnamefont {M.}~\bibnamefont
  {Nakatsuka}}, \bibinfo {author} {\bibfnamefont {N.}~\bibnamefont {Miyanaga}},
  \bibinfo {author} {\bibfnamefont {T.}~\bibnamefont {Kanabe}}, \bibinfo
  {author} {\bibfnamefont {H.}~\bibnamefont {Nakano}}, \ and\ \bibinfo {author}
  {\bibfnamefont {S.}~\bibnamefont {Nakai}},\ }\href@noop {} {\bibfield
  {journal} {\bibinfo  {journal} {Proc Spie}\ }\textbf {\bibinfo {volume}
  {1870}},\ \bibinfo {pages} {151} (\bibinfo {year} {1993})}\BibitemShut
  {NoStop}%
\bibitem [{\citenamefont {Webb}\ \emph {et~al.}(1992)\citenamefont {Webb},
  \citenamefont {Eimerl},\ and\ \citenamefont {Velsko}}]{Webb:92}%
  \BibitemOpen
  \bibfield  {author} {\bibinfo {author} {\bibfnamefont {M.~S.}\ \bibnamefont
  {Webb}}, \bibinfo {author} {\bibfnamefont {D.}~\bibnamefont {Eimerl}}, \ and\
  \bibinfo {author} {\bibfnamefont {S.~P.}\ \bibnamefont {Velsko}},\ }\href
  {\doibase 10.1364/JOSAB.9.001118} {\bibfield  {journal} {\bibinfo  {journal}
  {J. Opt. Soc. Am. B}\ }\textbf {\bibinfo {volume} {9}},\ \bibinfo {pages}
  {1118} (\bibinfo {year} {1992})}\BibitemShut {NoStop}%
\bibitem [{\citenamefont {Hillier}\ \emph {et~al.}(2013)\citenamefont
  {Hillier}, \citenamefont {Danson}, \citenamefont {Duffield}, \citenamefont
  {Egan}, \citenamefont {Elsmere}, \citenamefont {Girling}, \citenamefont
  {Harvey}, \citenamefont {Hopps}, \citenamefont {Norman}, \citenamefont
  {Parker}, \citenamefont {Treadwell}, \citenamefont {Winter},\ and\
  \citenamefont {Bett}}]{Hillier:13}%
  \BibitemOpen
  \bibfield  {author} {\bibinfo {author} {\bibfnamefont {D.}~\bibnamefont
  {Hillier}}, \bibinfo {author} {\bibfnamefont {C.}~\bibnamefont {Danson}},
  \bibinfo {author} {\bibfnamefont {S.}~\bibnamefont {Duffield}}, \bibinfo
  {author} {\bibfnamefont {D.}~\bibnamefont {Egan}}, \bibinfo {author}
  {\bibfnamefont {S.}~\bibnamefont {Elsmere}}, \bibinfo {author} {\bibfnamefont
  {M.}~\bibnamefont {Girling}}, \bibinfo {author} {\bibfnamefont
  {E.}~\bibnamefont {Harvey}}, \bibinfo {author} {\bibfnamefont
  {N.}~\bibnamefont {Hopps}}, \bibinfo {author} {\bibfnamefont
  {M.}~\bibnamefont {Norman}}, \bibinfo {author} {\bibfnamefont
  {S.}~\bibnamefont {Parker}}, \bibinfo {author} {\bibfnamefont
  {P.}~\bibnamefont {Treadwell}}, \bibinfo {author} {\bibfnamefont
  {D.}~\bibnamefont {Winter}}, \ and\ \bibinfo {author} {\bibfnamefont
  {T.}~\bibnamefont {Bett}},\ }\href {\doibase 10.1364/AO.52.004258} {\bibfield
   {journal} {\bibinfo  {journal} {Appl. Opt.}\ }\textbf {\bibinfo {volume}
  {52}},\ \bibinfo {pages} {4258} (\bibinfo {year} {2013})}\BibitemShut
  {NoStop}%
\bibitem [{\citenamefont {Zubairy}\ and\ \citenamefont
  {McIver}(1987)}]{PhysRevA.36.202}%
  \BibitemOpen
  \bibfield  {author} {\bibinfo {author} {\bibfnamefont {M.~S.}\ \bibnamefont
  {Zubairy}}\ and\ \bibinfo {author} {\bibfnamefont {J.~K.}\ \bibnamefont
  {McIver}},\ }\href {\doibase 10.1103/PhysRevA.36.202} {\bibfield  {journal}
  {\bibinfo  {journal} {Phys. Rev. A}\ }\textbf {\bibinfo {volume} {36}},\
  \bibinfo {pages} {202} (\bibinfo {year} {1987})}\BibitemShut {NoStop}%
\bibitem [{\citenamefont {Cai}\ and\ \citenamefont {Peschel}(2007)}]{Cai:07}%
  \BibitemOpen
  \bibfield  {author} {\bibinfo {author} {\bibfnamefont {Y.}~\bibnamefont
  {Cai}}\ and\ \bibinfo {author} {\bibfnamefont {U.}~\bibnamefont {Peschel}},\
  }\href {\doibase 10.1364/OE.15.015480} {\bibfield  {journal} {\bibinfo
  {journal} {Opt. Express}\ }\textbf {\bibinfo {volume} {15}},\ \bibinfo
  {pages} {15480} (\bibinfo {year} {2007})}\BibitemShut {NoStop}%
\bibitem [{\citenamefont {Dmitriev}\ \emph {et~al.}(2012)\citenamefont
  {Dmitriev}, \citenamefont {Osipov}, \citenamefont {Puzyrev}, \citenamefont
  {Sahakyan}, \citenamefont {Starodub},\ and\ \citenamefont
  {Vasin}}]{Dmitriev_2012}%
  \BibitemOpen
  \bibfield  {author} {\bibinfo {author} {\bibfnamefont {V.~G.}\ \bibnamefont
  {Dmitriev}}, \bibinfo {author} {\bibfnamefont {M.~V.}\ \bibnamefont
  {Osipov}}, \bibinfo {author} {\bibfnamefont {V.~N.}\ \bibnamefont {Puzyrev}},
  \bibinfo {author} {\bibfnamefont {A.~T.}\ \bibnamefont {Sahakyan}}, \bibinfo
  {author} {\bibfnamefont {A.~N.}\ \bibnamefont {Starodub}}, \ and\ \bibinfo
  {author} {\bibfnamefont {B.~L.}\ \bibnamefont {Vasin}},\ }\href {\doibase
  10.1088/0953-4075/45/16/165401} {\bibfield  {journal} {\bibinfo  {journal}
  {Journal of Physics B: Atomic, Molecular and Optical Physics}\ }\textbf
  {\bibinfo {volume} {45}},\ \bibinfo {pages} {165401} (\bibinfo {year}
  {2012})}\BibitemShut {NoStop}%
\bibitem [{\citenamefont {Vasin}\ \emph {et~al.}(2013)\citenamefont {Vasin},
  \citenamefont {B.}, \citenamefont {Korobkin}, \citenamefont {Yu.},
  \citenamefont {Osipov}, \citenamefont {M.}, \citenamefont {Puzyrev},
  \citenamefont {V.}, \citenamefont {Sahakyan},\ and\ \citenamefont
  {A.}}]{Vasin2013Second}%
  \BibitemOpen
  \bibfield  {author} {\bibinfo {author} {\bibnamefont {Vasin}}, \bibinfo
  {author} {\bibfnamefont {L.}~\bibnamefont {B.}}, \bibinfo {author}
  {\bibnamefont {Korobkin}}, \bibinfo {author} {\bibfnamefont {V.}~\bibnamefont
  {Yu.}}, \bibinfo {author} {\bibnamefont {Osipov}}, \bibinfo {author}
  {\bibfnamefont {V.}~\bibnamefont {M.}}, \bibinfo {author} {\bibnamefont
  {Puzyrev}}, \bibinfo {author} {\bibfnamefont {N.}~\bibnamefont {V.}},
  \bibinfo {author} {\bibnamefont {Sahakyan}}, \ and\ \bibinfo {author}
  {\bibfnamefont {T.}~\bibnamefont {A.}},\ }\href@noop {} {\bibfield  {journal}
  {\bibinfo  {journal} {Bulletin of the Lebedev Physics Institute}\ }\textbf
  {\bibinfo {volume} {40}},\ \bibinfo {pages} {205} (\bibinfo {year}
  {2013})}\BibitemShut {NoStop}%
  \bibitem [{\citenamefont {Cui}\ \emph {et~al.}()\citenamefont {Cui},
  \citenamefont {Gao}, \citenamefont {Rao}, \citenamefont {Liu}, \citenamefont
  {Li}, \citenamefont {Ji}, \citenamefont {Shi}, \citenamefont {Liu},
  \citenamefont {Zhao}, \citenamefont {Feng}, \citenamefont {Xia},
  \citenamefont {Liu}, \citenamefont {Li}, \citenamefont {Wang}, \citenamefont
  {Ma},\ and\ \citenamefont {Sui}}]{cui2019High}%
  \BibitemOpen
  \bibfield  {author} {\bibinfo {author} {\bibfnamefont {Y.}~\bibnamefont
  {Cui}}, \bibinfo {author} {\bibfnamefont {Y.}~\bibnamefont {Gao}}, \bibinfo
  {author} {\bibfnamefont {D.}~\bibnamefont {Rao}}, \bibinfo {author}
  {\bibfnamefont {D.}~\bibnamefont {Liu}}, \bibinfo {author} {\bibfnamefont
  {F.}~\bibnamefont {Li}}, \bibinfo {author} {\bibfnamefont {L.}~\bibnamefont
  {Ji}}, \bibinfo {author} {\bibfnamefont {H.}~\bibnamefont {Shi}}, \bibinfo
  {author} {\bibfnamefont {J.}~\bibnamefont {Liu}}, \bibinfo {author}
  {\bibfnamefont {X.}~\bibnamefont {Zhao}}, \bibinfo {author} {\bibfnamefont
  {W.}~\bibnamefont {Feng}}, \bibinfo {author} {\bibfnamefont {L.}~\bibnamefont
  {Xia}}, \bibinfo {author} {\bibfnamefont {J.}~\bibnamefont {Liu}}, \bibinfo
  {author} {\bibfnamefont {X.}~\bibnamefont {Li}}, \bibinfo {author}
  {\bibfnamefont {T.}~\bibnamefont {Wang}}, \bibinfo {author} {\bibfnamefont
  {W.}~\bibnamefont {Ma}}, \ and\ \bibinfo {author} {\bibfnamefont
  {Z.}~\bibnamefont {Sui}},\ }\href@noop {} {\bibinfo  {journal} {provisionally
  accepted by Optics Letters}\ }\BibitemShut {NoStop}%
\end{thebibliography}
%

\end{document}